\documentclass[%
 aip,
 amsmath,amssymb,
 reprint,%
]{revtex4-1}

\usepackage{ mathrsfs  }
\usepackage{graphicx}
\usepackage{dcolumn}
\usepackage{bm}
\usepackage{float}
\usepackage[dvipsnames]{xcolor}

\usepackage{gensymb}

\newcommand{\gpc}{\dot\gamma_{\rm c}}
\newcommand{\gpp}{\dot\gamma_{\rm p}}
\newcommand{\gp}{\dot\gamma}
\newcommand{\sigmap}{\sigma_{\rm p}}
\newcommand{\gy}{\gamma_{\rm y}}
\newcommand{\szero}{\sigma_0}
\newcommand{\sy}{\sigma_{\rm y}}
\newcommand{\Rcl}{R_\text{cl}}
\newcommand{\Rag}{R_\text{ag}}

\definecolor{colorr}{rgb}{0.85, 0, 0}
\definecolor{coloro}{rgb}{0.9, 0.4, 0.1}
\definecolor{colorb}{rgb}{0, 0.4470, 0.7410}
\definecolor{colorc}{rgb}{0.3010, 0.7450, 0.9330}
\definecolor{colorly}{rgb}{1.0000, 1.0000, 0.6250}
\definecolor{colorg}{rgb}{0.4660, 0.6740, 0.1880}
\definecolor{colorp}{rgb}{0.4940, 0.1840, 0.5560}
\definecolor{colorlr}{rgb}{1, 0.6, 0.6}
\definecolor{colorvdr}{rgb}{0.6, 0, 0}
\definecolor{color1}{rgb}{0, 1.0000, 0.3906}
\definecolor{color2}{rgb}{0, 0.9531, 1.0000}
\definecolor{color3}{rgb}{0, 0.3438, 1.0000}
\definecolor{color4}{rgb}{0.2188, 0, 1.0000}
\definecolor{color5}{rgb}{1.0000, 0, 0.7969}
\definecolor{color6}{rgb}{1.0000, 0.2188, 0}
\definecolor{color7}{rgb}{1.0000, 0.4271, 0}
\definecolor{color8}{rgb}{1.0000, 0.6875, 0}
\definecolor{color9}{rgb}{1.0000, 0.8438, 0}
\definecolor{color10}{rgb}{0,1,0.5}
\definecolor{color11}{rgb}{0,0.6667,0.6667}
\definecolor{color12}{rgb}{0,0.3294,0.8353}
\definecolor{color13}{rgb}{0,0,1}

\begin{document}

\title[Shear-induced memory effects in boehmite gels]{
Shear-induced memory effects in boehmite gels
}

\author{Iana Sudreau}
\affiliation{IFP Energies nouvelles, Rond-point de l’échangeur de Solaize, BP 3, 69360 Solaize, France\looseness=-1}
\affiliation{Univ Lyon, Ens de Lyon, Univ Claude Bernard, CNRS, Laboratoire de Physique, F-69342 Lyon,~France\looseness=-1}
\author{Sébastien Manneville}%
\affiliation{Univ Lyon, Ens de Lyon, Univ Claude Bernard, CNRS, Laboratoire de Physique, F-69342 Lyon,~France\looseness=-1}%
\author{Marion Servel}%
\affiliation{IFP Energies nouvelles, Rond-point de l’échangeur de Solaize, BP 3, 69360 Solaize, France\looseness=-1}
\author{Thibaut Divoux}%
\affiliation{Univ Lyon, Ens de Lyon, Univ Claude Bernard, CNRS, Laboratoire de Physique, F-69342 Lyon,~France\looseness=-1}%

\date{\today}

\begin{abstract}
Colloidal gels are formed by the aggregation of Brownian particles into clusters that are, in turn, part of a space-spanning percolated network. In practice, the microstructure of colloidal gels, which dictates their mechanical properties, strongly depends on the particle concentration and on the nature of their interactions. Yet another critical control parameter is the shear history experienced by the sample, which controls the size and density of the cluster population, via particle aggregation, cluster breakup and restructuring. Here we investigate the impact of shear history on acid-induced gels of boehmite, an aluminum oxide. We show that following a primary gelation, these gels display a dual response depending on the shear rate $\gpp$ used to rejuvenate their microstructure. We identify a critical shear rate $\gpc$, above which boehmite gels display a gel-like viscoelastic spectrum upon flow cessation, similar to that obtained following the primary gelation. However, upon flow cessation after shear rejuvenation below $\gpc$, boehmite gels display a glassy-like viscoelastic spectrum together with enhanced elastic properties. Moreover, the nonlinear rheological properties of boehmite gels also differ on both sides of $\gpc$: weak gels obtained after rejuvenation at $\gpp>\gpc$ show a yield strain that is constant, independent of $\gpp$, whereas strong gels obtained with $\gpp<\gpc$ display a yield strain that significantly increases with $\gpp$. Our results can be interpreted in light of previous literature on shear-induced {anisotropy}, which accounts for the reinforced elastic properties at $\gpp<\gpc$, while we rationalize the critical shear rate $\gpc$ in terms of a dimensionless quantity, the Mason number, comparing the ratio of the strength of the shear flow to the interparticle bond force. 
\end{abstract}

\maketitle

\section{Introduction}

Soft Glassy Materials (SGM) are ubiquitous in our everyday life. They encompass a broad range of viscoelastic materials that are key to out-of-equilibrium living organisms and omnipresent in major industries, e.g.,~foodstuff, personal care, and building industry \cite{Gibaud:2012,Ioannidou:2016,Spicer:2020}. SGMs are composed of sub-units such as particles or polymers, whose interactions and volume fraction control the macroscopic mechanical properties of these soft materials \cite{Sciortino:2002,Bonnecaze:2010,RuizFranco:2020}. In practice, SGMs display solid-like properties at rest and under small strains, whereas they yield and flow like liquids when submitted to a large enough strain \cite{Barnes:1999,Bonn:2017}. 
At rest, their solid-like behavior originates either from the presence of attractive interactions between constituents, which form a space-spanning network at low volume fractions, or from the jamming of the constituents for large enough volume fractions. The former category of SGMs is referred to as gels, whereas the latter category is referred to as soft glasses \cite{,Zaccarelli:2009,Lu:2013,Bonn:2017}.

Beside volume fraction and particle interactions, an additional control parameter of SGM properties at rest is the route followed towards the solid-like state through the shear history experienced during the liquid-to-solid transition. In soft glasses, oscillatory shear deformation of moderate amplitude can be used to mechanically encode some memory into the material through local plastic deformations \cite{Fiocco:2014,Lavrentovich:2017,Keim:2019}. The imprint of the shear protocol may subsequently be read out by a strain sweep from small to large amplitudes \cite{Mukherji:2019}. 
Similar memory effects have been identified in colloidal gels \cite{Schwen:2020}, where shear history significantly affects the microstructural and mechanical properties \cite{Altmann:2004,Koumakis:2015,Moghimi:2017}. Indeed, depending on its intensity, mechanical shear may either enhance or compete with the attractive interactions driving the formation of clusters, which play a key role in the solid-like behavior of the sample upon flow cessation \cite{Zaccone:2009,Whitaker:2019}. Moreover, upon partial yielding, gels often break down in a spatially anisotropic way \cite{Varadan:2001,Hoekstra:2005,Rajaram:2010,Masschaele:2011,Landrum:2016,Boromand:2017}, and such anisotropy gets frozen into the gel microstructure during flow cessation \cite{Colombo:2017,Moghimi:2017b}. In that framework, shear history strongly impacts the rheological properties of colloidal gels, both in their fluidized state as weakly aggregated suspensions and upon flow cessation, when they {reform  into a viscoelastic solid. The present article is devoted to the rheological fate of a colloidal gel composed of boehmite particles after shear is applied with a constant rate. In the rest of this introduction, we briefly recall the phenomenology of thixotropy in colloidal gels and how their sensitivity to shear can be used to induce ``memory'' effects and tune their microstructure. We then present the state-of-the-art specific to boehmite gels and summarize our main results.}

{When a shear stress much larger than the yield stress is applied to a colloidal gel, its} microstructure is broken down into clusters of particles with attractive interactions. In such a suspension, the balance between hydrodynamic forces and internal cohesive forces determines whether shear induces aggregation, break-up or internal restructuring of the cluster \cite{Harshe:2011,Bubakova:2013}, among which the latter two phenomena are thermally activated \cite{Conchuir:2013}. Starting from a fully dispersed state, i.e., from a suspension of individual particles, increasing the shear {rate} accelerates the growth and decreases the size of clusters, whose steady-state properties depend on both the shear {rate} and the particle concentration \cite{Oles:1992,Serra:1997}. More generally, for any configuration in the fluidized state, i.e., partially or fully dispersed, the gel response during transient flows depends on whether the shear is increasing or decreasing. This effect leads to thixotropy and rheological hysteresis, which are ubiquitously observed when measuring a constitutive equation, shear stress $\sigma$ vs. shear rate $\dot \gamma$ \cite{Mewis:2009,Larson:2019}. Such hysteresis loops have been characterized through their area, which is maximum for a critical sweep rate of the shear, pointing towards a thixotropic timescale that is characteristic of the sample \cite{Divoux:2013,Radhakrishnan:2017,Jamali:2019}. {Thixotropy thus relates to the sample ``memory'' of the past shear history.} 

{Another manifestation of shear-induced memory in colloidal gels occurs upon flow cessation. Indeed, it is now well-established that the microstructure of the gel that rebuilds once shear is stopped is strongly affected by the value of the previously applied shear rate. For instance, Koumakis {\it et al.}\cite{Koumakis:2015} have shown that} high shear rates fully break the structure of {depletion gels of poly(methyl methacrylate) colloids} and lead, after shear cessation, to more homogeneous and stronger gels. In contrast, preshear at low shear rates creates largely heterogeneous and much weaker gels with reduced elasticity \cite{Koumakis:2015}. Such sensitivity to shear history has often been presented as a limitation that needs to be overcome to {accurately control the properties of colloidal gels}. However, shear history can be used as a way to {finely tune} the liquid-to-solid transition and the resulting microstructure of the gel formed {after} flow cessation\cite{Moghimi:2017}. 
In that spirit, shear-assisted gelation of colloidal suspensions of carbon black particles was controlled through the rate of flow cessation, which allows tuning the microstructure of the gel continuously \cite{Helal:2016}. Fast flow cessations lead to strong gels with a highly connected microstructure, whereas slow flow cessations produce weaker gels composed of poorly connected aggregates. These results were further confirmed by means of a rheo-impedance study on more complex mixtures, e.g., carbon black suspensions in a semisolid flow batteries solvent \cite{Narayanan:2017}, and more recently on carbon black suspensions in propylene carbonate through shear rheology coupled with small angle neutron scattering \cite{Hipp:2019}.

{To summarize, the microstructure of colloidal gels is exquisitely sensitive to shear history. This leads to shear-induced memory not only through reversible, thixotropic effects under flow, but also through permanent modifications of the microstructure after flow cessation. The present work is concerned with the latter type of memory effects, which we also identify with the ``shear-induced tunability'' of colloidal gels recently introduced in the literature\cite{Koumakis:2015,Moghimi:2017}.} In the case of model spherical particles with short-range attractive interactions, the {interplay between shear and microstructure} can be rationalized in terms of a dimensionless ratio, called the Mason number Mn, comparing the strength of the shear flow to the interparticle bond force at contact \cite{Markutsya:2014,Boromand:2017,Varga:2018,Varga:2019}. At low Mason number, ${\rm Mn}< 10^{-2}$, a gel behaves as a viscoelastic solid. For intermediate values, $10^{-2}\leq {\rm Mn} \leq 1$, constant breakup and reformation of interparticle bonds results in a complex heterogeneous dynamics, whereas for ${\rm Mn}>1$ the gel is fluidized into a weakly aggregated suspension of clusters and particles \cite{Jamali:2020}. Yet, this framework, which holds for model spherical particles, might not account for particles with complex shape and interactions, including chemical reactions and irreversible binding, thus calling for further experimental investigations of non-spherical colloidal gels of industrial interest. 

In the present paper, we study the impact of shear history on gels of boehmite, an aluminum oxyhydroxide (AlO(OH)), which is a precursor of gamma alumina ($\gamma$--Al$_{2}$O$_{3}$) widely used in industries to design catalyst supports  {\cite{Euzen:2002,Xiong:2014,Zheng.2014}}. These gels are part of a large family of systems, whose primary sol-gel transition involves chemical reactions, which result in the formation of strong aggregates that are unbreakable, even when submitted to extended period of high shear {rate}. Note that aside boehmite gels, other examples include dispersions of silica colloids in presence of {a sufficient amount of} salt, whose addition triggers the fast and irreversible aggregation of the colloids through the formation of permanent interparticle siloxane bonds \cite{Depasse:1970,Depasse:1997,Kurokawa:2015}.
These systems are thus composed of complex building blocks most of which are unbreakable aggregates of the original colloids. Following this peculiar ``primary'' sol-gel transition, the properties of boehmite gels are stable in time, and cycles of shear-induced fluidization followed by gelation upon flow cessation can be consistently repeated. Such ``secondary" sol-gel transitions are the topic of the present study in which we focus on the impact of the shear {rate} applied during the shear-rejuvenation step. 

In practice, starting from a sample that has experienced a primary sol-gel transition in quiescent conditions, we impose a shear-rejuvenation step at a shear rate $\gpp$, following which the flow is stopped  and the sample left at rest. The linear and nonlinear viscoelastic properties of the gel that eventually reforms are monitored before imposing a new shear-rejuvenation step with a different shear {rate}. We explore three decades in values of $\gpp$ to determine the impact of shear history on boehmite gels over a broad range of Mason numbers.
Following this protocol, we show that after a primary gelation, boehmite gels can be rejuvenated reversibly by an external shear, whose intensity strongly impacts the subsequent properties of the gels that reform following flow cessation. In particular, we identify a critical shear rate $\gpc$, above which boehmite gels display a gel-like viscoelastic response upon flow cessation, similar to that obtained following the primary gelation. However, for shear rejuvenation such that $\gpp<\gpc$, boehmite gels display upon flow cessation a glassy-like viscoelastic spectrum together with enhanced elastic properties. The value of $\gpc$ is found to be independent of both the boehmite and acid concentration over the range explored.

In light of the framework built on the Mason number, we propose an interpretation of our observations in which gels sheared at $\gpp>\gpc$ are fully rejuvenated and fluidized into individual unbreakable aggregates, which yields a gel upon flow cessation that shows similar mechanical and microstructural properties to that obtained after a primary gelation. On the contrary, shearing gels at $\gpp<\gpc$ results in the shear-induced formation of marginally stable clusters of larger dimensions than in the primary gel, {and of anisotropic shape, leading to enhanced elastic properties}. The latter scenario, in which the gel is not fully rejuvenated but instead encodes some memory of the flow history into its microstructure, is a prototypical example of what was recently coined ``directed aging" in the literature \cite{Pashine:2019,Hexner:2020}. Finally, we demonstrate that imposing stress ramps, instead of abrupt flow cessation, allows us to continuously tune the linear and non-linear rheological properties of boehmite gels between the two extreme cases discussed above. Our results show that shear history is a promising candidate for tailoring the microstructure of catalyst supports obtained from these soft precursors. {They could also be relevant to a wider range of colloids as they do not seem to depend on the specific aggregation kinetics that initiate the gelation.}

\section{Materials and methods}

\subsection{Preparation of boehmite gels}

\begin{figure}[t!]
\includegraphics[scale=0.415]{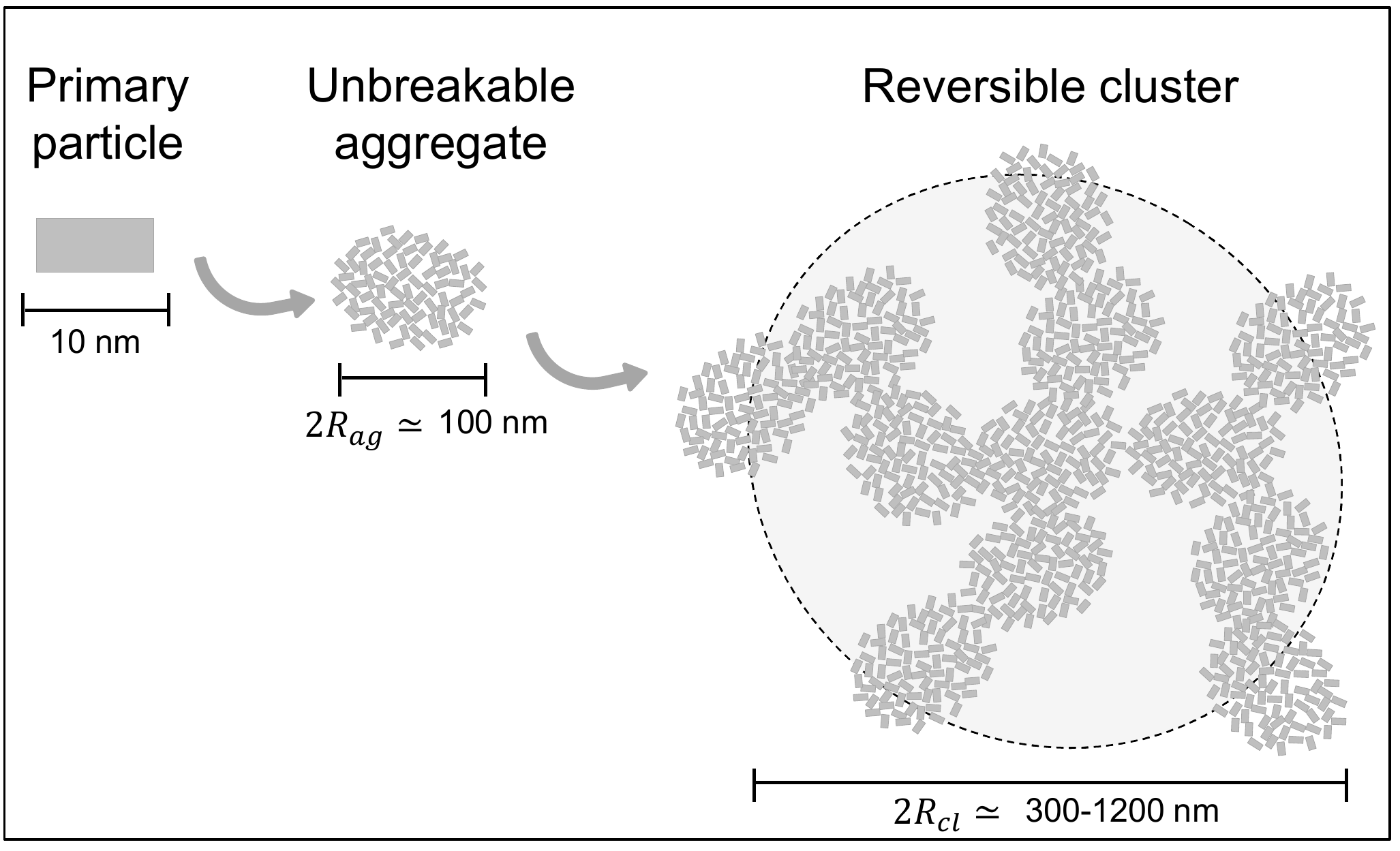}
\caption{\label{primary-structure} Sketch of the structure of a reversible cluster (highlighted by a light grey disk) composed of several unbreakable aggregates of primary particles and formed during the primary sol-gel transition.} 
\end{figure}

Boehmite gels are prepared by adding a boehmite powder (Plural SB3, Sasol) to an aqueous solution of nitric acid \cite{Ramsay.1978,Drouin.1988,Cristiani.2007}. Unless specified otherwise, the boehmite and acid concentrations are respectively 123~g.L$^{-1}$ and 14~g.L$^{-1}$, {which corresponds to a solid volume fraction of 4$\%$ in boehmite}. These concentrations were chosen to produce {dilute} gels, whose secondary gelation takes place within a few seconds after flow cessation.
In practice, the boehmite powder is dispersed into a solution of nitric acid, first by mixing at 600 rpm during 20 min before mixing at 1800 rpm during 15 min. The result is a suspension containing {anisotropic {\it primary particles} with a typical size of about 10~nm}{\cite{Gallois.,Fauchadour.2002,Speyer:2020,Zheng.2014} (see sketch in Fig.~\ref{primary-structure}). This suspension} is left at rest for at least one week prior to any rheological test. Indeed, while the primary sol-gel transition occurs after a couple of hours, it takes several days for the pH of the suspensions to stabilize at $\text{pH}=3.5$ due to the dissolution and the surface charging of the alumina \cite{Cristiani.2007,Fauchadour.2002}. The presence of the nitrate anion NO$_{3}^{-}$ screens the electrostatic repulsive interactions between the positively charged {surfaces of the boehmite particles \cite{Wood:1990,Speyer:2020,Raybaud.2001,Zheng.2014}, which} eventually self-assemble into {{\it unbreakable} aggregates} of diameter ${2}\Rag\simeq {100}$~nm {as determined from light scattering measurements on sheared dilutions of primary gels \cite{Sudreau:InPrep}, and in good agreement with previous cryo-TEM measurements on similar boehmite dispersions \cite{Fauchadour.2002}. These aggregates are subject to short-range van der Waals attractive interactions and}, as sketched in Fig.~\ref{primary-structure}, further assemble into a percolated network of large clusters of typical diameter ${2} \Rcl\simeq 300$--1200~nm as determined from light scattering measurements \cite{Sudreau:InPrep}. The elastic properties of these gels are dominated by the fractal nature of the clusters, which results in a power-law increase of the gel storage modulus with the boehmite content \cite{Shih:1990,Drouin.1988}. {Finally, under external shear, these gels display a strong thixotropic behaviour as evidenced by broad hysteresis loops up to large shear rates \cite{Song:1989} [see Fig. \ref{fig:flow_curve_cisaillement} in Appendix~\ref{Appendix_flow_curve}].} 

\subsection{Experimental set-up and protocol}

Rheological measurements are performed in a smooth cylindrical Couette geometry (height 58~mm, rotating inner cylinder of radius 24~mm, fixed outer cylinder of radius 25~mm, gap $e=1$~mm). The cell is topped with a homemade lid to minimize evaporation, and the rotor is connected to a stress-controlled rheometer (AR-G2, TA Instruments). The Couette cell is immersed in a water bath, allowing all the experiments to be performed at a fixed temperature of ($20.0\pm 0.1)~\degree$C.

First, we determine the rheological properties of a pristine or ``primary'' gel, which forms for the first time in quiescent conditions, i.e., in the absence of any external shear. In practice, a fresh sample is loaded in the Couette cell right after its preparation and submitted to the three following steps : ($i$) a period of rest of 3000~s during which the linear viscoelastic properties are monitored under small-amplitude oscillatory shear ($f= 1$~Hz, $\gamma = 0.1 \%$),  followed by ($ii$) a frequency sweep of total duration $156$~s from $f=10$~Hz to $f=0.1$~Hz at $\gamma = 0.1 \%$. Finally, we perform ($iii$) a strain amplitude sweep of total duration $275$~s from $\gamma=0.01$~\% to $\gamma=100$~\% at $f= 1$~Hz to determine the nonlinear response and the yield strain of the sample.

Following the primary sol-gel transition, the properties of the samples are stable in time, provided that the sample does not suffer from evaporation, i.e., typically over 12~h with our experimental setup. The core of the present study is performed on non-pristine samples, i.e., gels obtained by ``secondary'' gelations, which experience multiple sequences of shear and rest periods following their primary gelation. In practice, a large batch of about 75~mL of pristine gel is prepared and kept at rest in a closed container of volume 100~mL for at least seven days. For a given series of measurements, 7.5~mL of {this pristine} gel is transferred into the Couette cell. {The} sample is then rejuvenated by shearing at a constant shear rate $\gpp$ during 600 s. After shear rejuvenation, the secondary gelation process is characterized by the exact same three-step protocol ($i$), ($ii$), and ($iii$) described above. Such shear rejuvenation and secondary gelation are successively repeated for nine values of shear rates $\gpp$ ranging from 2~s$^{-1}$ to 1000~s$^{-1}$. {Since the experiments are performed on the same sample, the values of $\gpp$ are chosen in a random order to make sure that the elastic modulus obtained afterwards is related to $\gpp$ and not to some aging phenomena associated with accumulated strain.}

\section{Experimental results}

\subsection{Impact of the shear-rejuvenation step on the gel linear viscoelastic properties}
\label{LinearVisco}

\begin{figure}[t!]
\includegraphics[scale=0.55]{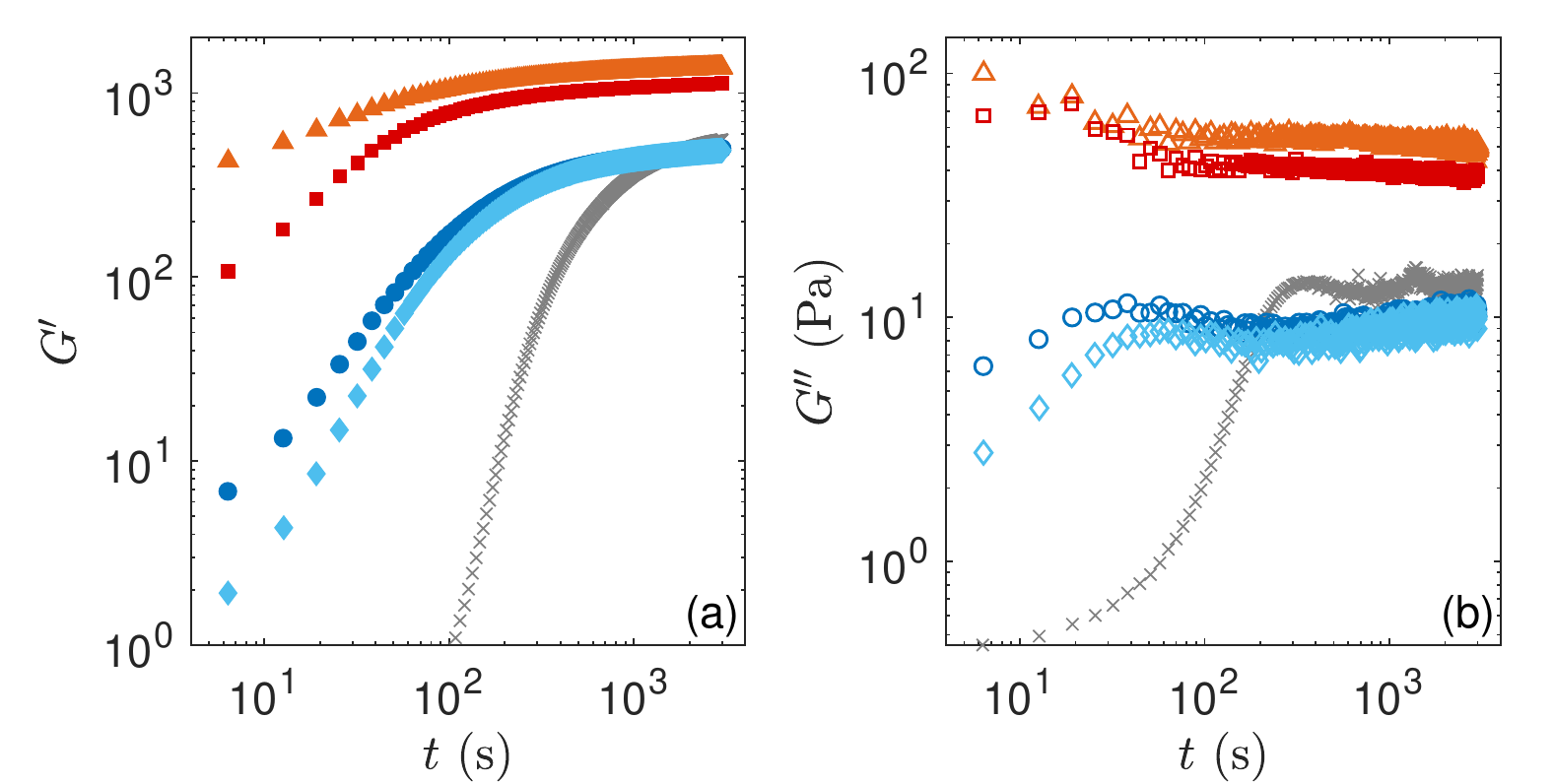}
\caption{\label{reprise}Temporal evolution of (a) the storage modulus $G'$ and (b) the loss modulus $G''$ for a primary gelation (\textcolor{gray}{ $\times$}), and for subsequent secondary gelations all preceded by a 600~s shear rejuvenation under various shear rates $\gpp=2$~s$^{-1}$ (\textcolor{coloro}{$\blacktriangle$}), 10~s$^{-1}$ (\textcolor{colorr}{$\blacksquare$}), 100~s$^{-1}$ (\textcolor{colorb}{$\bullet$}), 500~s$^{-1}$ (\textcolor{colorc}{$\blacklozenge$}). For secondary gelations, the origin of time is taken at the end of the shear-rejuvenation step. Viscoelastic properties are measured under small-amplitude oscillatory shear ($f=1$~Hz, $\gamma=0.1$~\%).} 
\end{figure}

Figure~\ref{reprise} shows the temporal evolution of the gel linear viscoelastic properties during the primary gelation, and for a series of four subsequent secondary gelations, each preceded by a 600~s preshear of constant intensity $\gpp=2, 10, 100$~s$^{-1}$ and 500~s$^{-1}$. During the primary gelation, the formation of a space-spanning colloidal network, defined as a first estimate by the crossover point between $G'$ and $G''$, occurs after about $t \simeq 100$~s. The gelation dynamics are noticeably faster for the subsequent gelations, where the crossover takes place within the first 20~s, independently of the shear rate $\gpp$ applied during the rejuvenation step. Moreover, for all the secondary gelations, the storage modulus $G'$ increases over time and reaches a plateau at $t \simeq 1000$~s, which value $G'_0$ depends on the applied shear rate $\gpp$ during the rejuvenation step. A first important result is that the terminal values of $G'_0$ and $G''_0$ reached at the end of a secondary gelation following strong shear rejuvenation (i.e., $\gpp=100$~s$^{-1}$ and 500~s$^{-1}$) are comparable with those obtained after the primary gelation. On the contrary, gels obtained after shear rejuvenation at a low intensity show stronger viscoelastic properties, by a factor of about 2.5 for the storage modulus and 4 for the loss modulus. 

\begin{figure*}[!]
\includegraphics[scale=0.65]{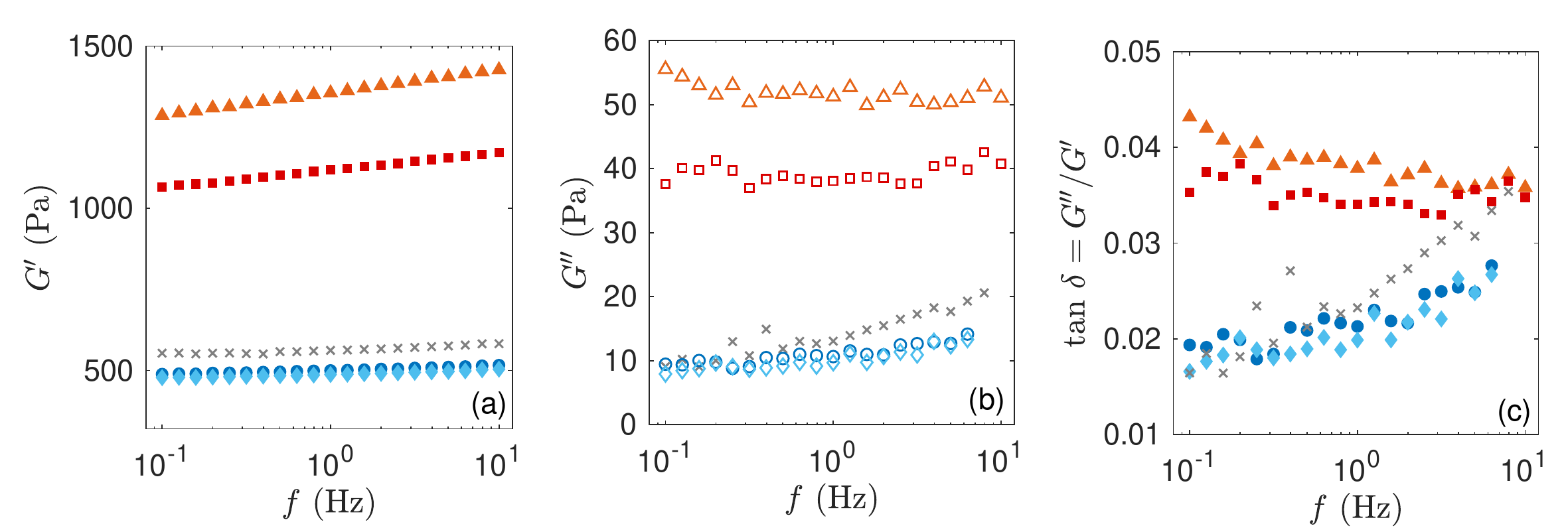}
\caption{\label{frequency_sweep} Frequency dependence of (a) the storage modulus $G'$, (b) the loss modulus $G''$, and (c) the loss factor $\tan \delta$ following a primary gelation of 3000~s (\textcolor{gray}{ $\times$}) or a secondary gelation including a rest period of 3000~s preceded by a 600~s shear-rejuvenation step under a shear rate $\gpp=2$~s$^{-1}$ (\textcolor{coloro}{$\blacktriangle$}), 10~s$^{-1}$ (\textcolor{colorr}{$\blacksquare$}), 100~s$^{-1}$ (\textcolor{colorb}{$\bullet$}), 500~s$^{-1}$ (\textcolor{colorc}{$\blacklozenge$}). Viscoelastic spectra measured with a strain amplitude $\gamma=0.1$~\%.}
\end{figure*}

The terminal state of the four secondary gelations are further characterized by measuring the linear viscoelastic spectrum over two decades of frequency. For both the primary and secondary gelations, the storage modulus shows a logarithmic increase with the frequency {[Fig.~\ref{frequency_sweep}(a)]}, whose slope is larger in the case of gels formed after a rejuvenation step of low shear {rate}, whereas the loss modulus hardly depends on $f$ {[Fig.~\ref{frequency_sweep}(b)]}. Concomitantly, the loss factor $\tan\delta$ allows us to split the data into two groups: the loss factor is an increasing function of the frequency for the primary gel and secondary gels formed after a rejuvenation step of high shear {rate}, whereas the loss factor is a decreasing function of $f$ for secondary gels formed after a rejuvenation step of low shear {rate} [Fig.~\ref{frequency_sweep}(c)]. This second remarkable result suggests that gels formed after a low shear {rate} display a different microstructure than the primary gel and secondary gels formed after a rejuvenation step of high shear {rate}.

\begin{figure}[!t]
\includegraphics[scale=0.65]{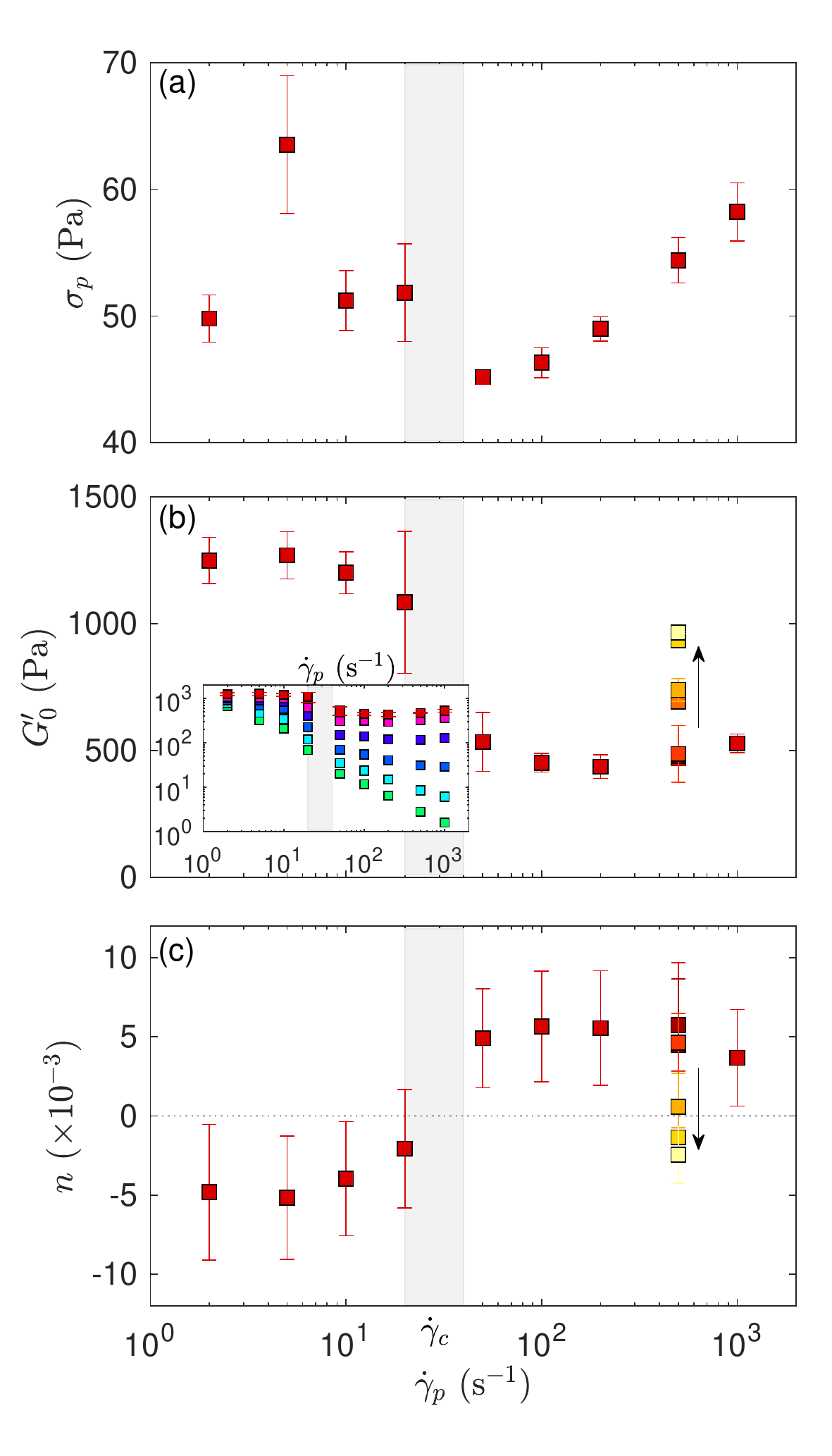}
\caption{\label{G_gamma}Dependence on the shear rate $\gpp$ used for shear rejuvenation of (a) the stress $\sigma_{p}$ measured at the end of the shear-rejuvenation step, (b) the storage modulus $G'$ measured after 3000~s of rest period following the shear-rejuvenation step, and (c) the slope $n$ of the linear regression of the loss factor $\tan\delta$ versus $\log f$.
The shear-rejuvenation step is either stopped abruptly ($\dot \gamma=0$~s$^{-1}$, red symbols) or stopped over a finite duration $\Delta t$ using a linear decreasing ramp of stress following a 600~s shear rejuvenation at $\gpp=500$~s$^{-1}$ (other colored symbols). The duration $\Delta t$ of the ramp is $\Delta t= 0$~s (\textcolor{colorr}{$\blacksquare$}), $\Delta t= 54$~s (\textcolor{color6}{$\blacksquare$}), $\Delta t= 270$~s (\textcolor{color7}{$\blacksquare$}), $\Delta t= 540$~s (\textcolor{color8}{$\blacksquare$}), $\Delta t= 3240$~s (\textcolor{color9}{$\blacksquare$}) and $\Delta t= 12 960$~s (\textcolor{colorly}{$\blacksquare$}). The inset in (b) shows the storage modulus $G'$ vs $\gpp$ measured at different points in time $t$ following the abrupt cessation of shear rejuvenation [$t=6$~s (\textcolor{color1}{$\blacksquare$}), $13$~s~ (\textcolor{color2}{$\blacksquare$}), $26$~s (\textcolor{color3}{$\blacksquare$}), $63$~s
(\textcolor{color4}{$\blacksquare$}), $300$~s (\textcolor{color5}{$\blacksquare$}), and $3000$~s (\textcolor{colorr}{$\blacksquare$})]. Error bars in (a) and (b) represent one standard deviation computed on three to eight independent tests. Error bars in (c) represent the root-mean-square error of the linear regression. The grey rectangle highlights the critical shear rate $\gpc=30\pm$10~s$^{-1}$.}
\end{figure}

To get a better sense of the impact of the shear {rate} applied during the rejuvenation step, we have systematically explored the impact of $\gpp$ applied during the shear-rejuvenation step preceding secondary gelations. The results are gathered in Fig.~\ref{G_gamma}, where we report over three decades of $\gpp$, the stress $\sigma_p$ at the end of the shear-rejuvenation step, the terminal value of the gel storage modulus $G'_0$ determined 3000~s after the end of the rejuvenation step, and the slope $n$ characterizing the logarithmic dependence of the loss factor with $f$, i.e., $G' \sim n\log f$. These three observables exhibit two distinct behaviors separated by a critical shear rate~$\dot{\gamma}_{c}= 30 \pm$10~s$^{-1}$. More specifically, the stress at the end of the shear-rejuvenation step $\sigma_p$ {does not exhibit a clear dependence on} $\gpp$ for $\gpp<\gpc$, whereas $\sigma_p$ increases logarithmically with $\gpp$ for $\gpp>\gpc$ [Fig.~\ref{G_gamma}(a)]. Moreover, secondary gelations following a rejuvenation step such that $\gpp<\gpc$ lead to stronger gels, with $G'_0=1250\pm150$~Pa, whereas secondary gelations following a rejuvenation step such that $\gpp>\gpc$ lead to gels of  elasticity comparable to that of a primary gel, i.e., $G'_0=500\pm150$~Pa vs $G'_0=580$~Pa, respectively [Fig.~\ref{G_gamma}(b)]. Remarkably, the storage modulus is roughly independent of the shear {rate} on both sides of $\gpc$. Finally, a similar binary outcome is found on the slope $n$ of the logarithmic scaling of the loss factor $\tan \delta$ with frequency [Fig.~\ref{G_gamma}(c)]. For $\gpp<\gpc$, $n$ is negative, whereas for $\gpp>\gpc$, $n$ is positive similarly to what is observed for a primary gelation [Fig.~\ref{frequency_sweep}(c)]. This last result shows that on both sides of $\gpc$, the gels display opposite relaxation behaviors \cite{Winter:2013}, which points to significant differences in the microstructure of gels reformed after shear rejuvenation above or below $\gpc$.  

\begin{figure}
\includegraphics[scale=0.65]{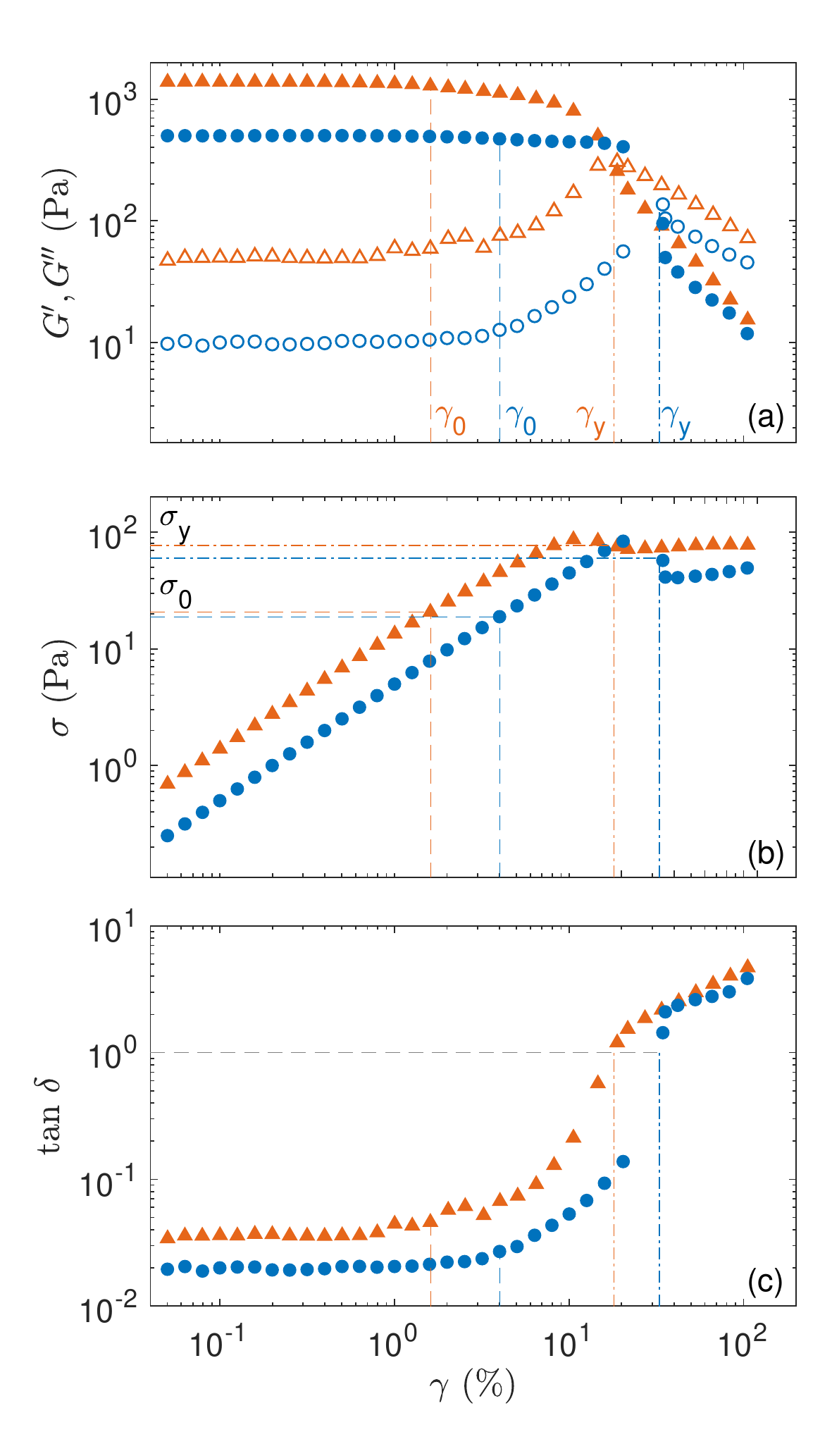}
\caption{\label{amplitude_sweep} Strain dependence of (a) the storage modulus $G'$ (full symbols) and the loss modulus $G''$ (empty symbols), (b) the stress amplitude $\sigma$, and (c) the loss factor tan $\delta$ of a boehmite gel obtained in a secondary gelation following a rejuvenation step of intensity: $\gpp=2$~s$^{-1}$ (\textcolor{coloro}{$\blacktriangle$}) and 500~s$^{-1}$ (\textcolor{colorb}{$\bullet$}). Strain sweep experiment performed at $f=1$~Hz. In (a)--(c), the vertical dashed and dashed-dotted lines respectively mark the strain $\gamma_0$ beyond which the gel response becomes nonlinear and the yield strain $\gy$ defined as the crossover of $G'$ and $G''$. }
\end{figure}

\subsection{Impact of the shear-rejuvenation step on the gel nonlinear viscoelastic properties}
\label{NonlinearVisco}

We now turn to the impact of the shear {rate} $\gpp$ on the nonlinear response of boehmite gels and on their yielding scenario. Figure~\ref{amplitude_sweep}(a) shows the evolution of $G'$ and $G''$ during a strain sweep for gels obtained after two rejuvenation steps performed at low and high shear {rate}, respectively $\gpp=2$~s$^{-1}$ and 500~s$^{-1}$. At low strain amplitude, i.e., $\gamma \lesssim 1$~\%, both gels display a strain-independent response, which is mainly elastic, i.e., such that $G'_0\gg G''_0$. Upon increasing the strain amplitude $\gamma$, the storage modulus decreases monotonically and departs from the linear regime at a strain $\gamma_0$ defined arbitrarily as $G'(\gamma_0)= 0.95G'_0$. Concomitantly, the loss modulus increases and shows a maximum in the vicinity of the yield strain $\gy$, defined as the locus of the crossover between $G'$ and $G''$, before decreasing for $\gamma>\gy$. Throughout the linear regime and above $\gamma_0$, the amplitude $\sigma$ of the oscillatory stress response increases up to the yield point, beyond which it becomes roughly strain independent [Fig.~\ref{amplitude_sweep}(b)]. Finally, the loss factor shows a similar monotonic trend for both gels, which is indicative of the transition from a solid-like response to a liquid-like one beyond $\gy$ [Fig.~\ref{amplitude_sweep}(c)]. To summarize, the monotonic decrease of $G'(\gamma)$ and the non-monotonic shape of $G''(\gamma)$, which are prototypical of the nonlinear response of numerous yield stress fluids including colloidal gels \cite{Gibaud:2009,Gibaud:2010,Grenard:2014} and glasses \cite{Mason:1995b}, dense emulsions, and foams \cite{Mason:1995c,Cohen:2014}, are qualitatively insensitive to the shear {rate} $\gpp$ of the shear-rejuvenation step.    

\begin{figure*}
\includegraphics[scale=0.65]{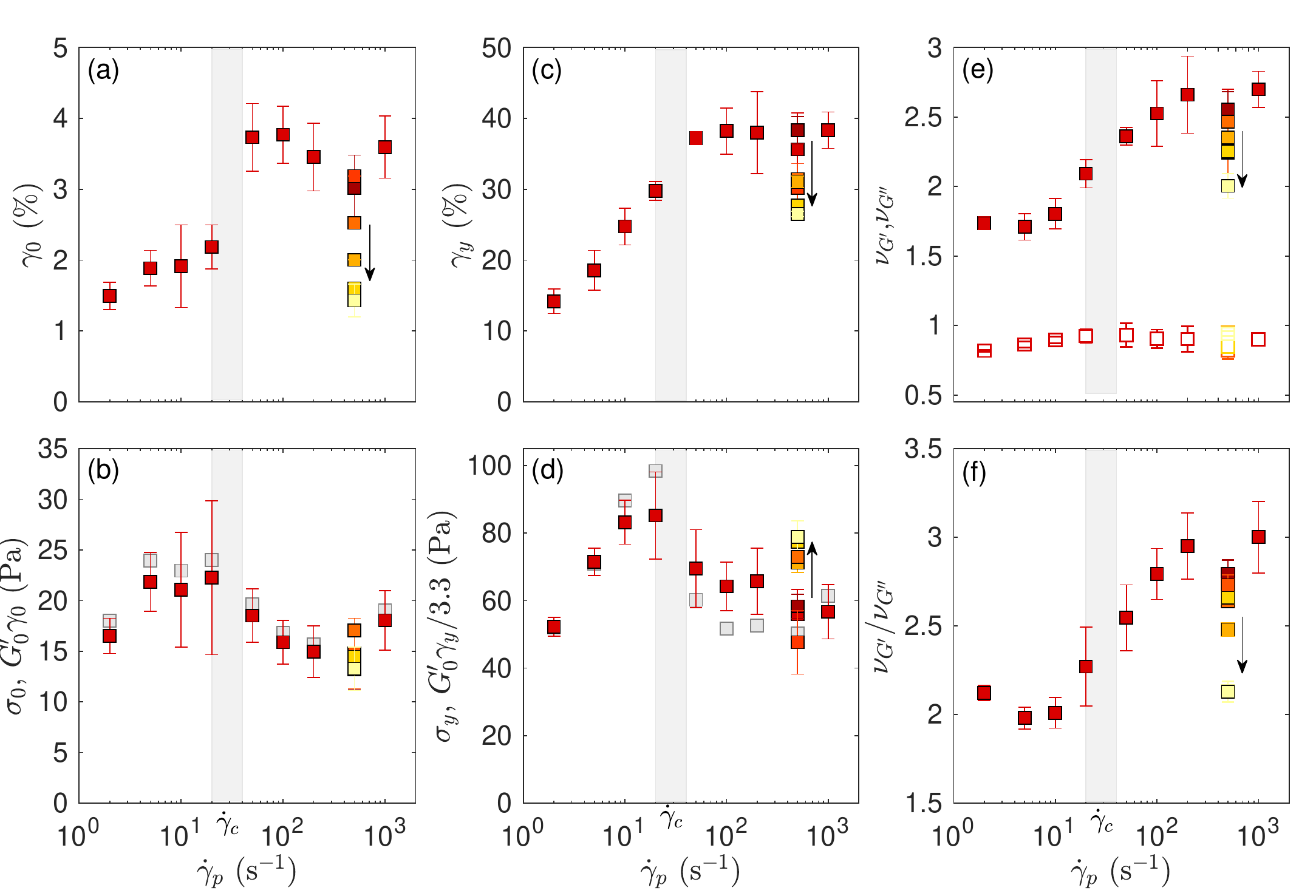}
\caption{\label{sigma_gamma}  (a) Strain amplitude $\gamma_{0}$ at the onset of the nonlinear regime and (b) corresponding stress amplitude $\sigma_{0}$ 
vs the shear rate $\gpp$ applied during the shear-rejuvenation step. (c) Yield strain $\gy$ and (d) yield stress $\sy$ vs $\gpp$. 
(e) Exponent $\nu_{G'}$ (filled symbols) and $\nu_{G''}$ (open symbols) characterizing the power-law decrease of $G'$ and $G''$ beyond the yield point vs $\gpp$ and (f) ratio $\nu_{G'}/\nu_{G''}$ vs $\gpp$. Grey squares in (b) and (d) respectively show $G^{\prime}_{0} \gamma_{0}$ and $G^{\prime}_{0} \gamma_{y}/3.3$. {The scaling factor 3.3 has been chosen to allow for a direct comparison of the evolution of $G'_0\gamma_y$ with that of $\sy$.} Same symbols, color code, and error bar estimations as in Fig.~\ref{G_gamma}.} 
\end{figure*}

We note, however, that the strain $\gamma_0$ that marks the end of the linear regime and the yield strain $\gy$ {show a similar evolution} for the whole range of shear {rates} $\gpp$ under study, as illustrated in Figs.~\ref{sigma_gamma}(a) and \ref{sigma_gamma}(c). More precisely, secondary gelations following shear rejuvenation at $\gpp<\gpc$ yield gels where $\gamma_0$ and $\gy$ increase for increasing $\gpp$, whereas $\gpp>\gpc$ leads to gels with constant strain values, i.e,  $\gamma_0 \simeq 3.5$\% and $\gy \simeq 38$\%, independent of $\gpp$. These values are comparable to those measured following a primary gelation, namely $\gamma_0 \simeq 5$\% and $\gy \simeq 29$\%. 

We now focus on the stress $\szero$ at the onset of the nonlinear regime and the yield stress $\sy$, which respectively correspond to the amplitudes of the stress responses measured for strain amplitudes $\gamma_0$ and $\gy$ [see Fig.~\ref{amplitude_sweep}(b)]. The dependence of $\szero$ and $\sy$ with the shear rate $\gpp$ applied during the shear-rejuvenation step is reported in Figs.~\ref{sigma_gamma}(b) and \ref{sigma_gamma}(d). Except perhaps for a weak maximum at $\gpp\simeq\gpc$, the two characteristic stresses $\szero$ and $\sy$ do not show any clear systematic trend with $\gpp$. Note that the evolution of $\szero$ and $\sy$ are very similar to those of $G'_0\gamma_0$ and $G'_0\gy$, respectively [see grey squares in Figs.~\ref{sigma_gamma}(b) and \ref{sigma_gamma}(d)], which suggests that $\szero$ and $\sy$ are essentially governed by the product of the elastic modulus and the corresponding characteristic strain. {The fact that the stress mainly coincides with the product of the elastic modulus and the strain at the yield point suggests that the yielding transition, which {can be} described as a continuous transition from recoverable to non-recoverable strains{\cite{Petekidis.2003,Donley.2020}}, is here rather abrupt.}

Finally, beyond the yield point, the decrease of the storage and loss modulus are both well fitted by power-law decays $G' \sim \gamma^{-\nu_{G'}}$ and $G'' \sim \gamma^{-\nu_{G''}}$ [see Fig.~\ref{supfig2} in Appendix~\ref{Appendix2}]. The dependence of the exponents $\nu_{G'}$ and $\nu_{G''}$ with $\gpp$ is reported in Fig.~\ref{sigma_gamma}(e), while that of the ratio $\nu_{G'}/\nu_{G''}$ is reported in Fig.~\ref{sigma_gamma}(f). The decrease in $G'$ beyond the yield point is all the more steep than $\gpp$ is large, i.e., $\nu_{G'}$ increases with $\gpp$, whereas the corresponding exponent for $G''$ is poorly sensitive to the shear {rate} $\gpp$. Interestingly, for $\gpp<\gpc$, the ratio  $\nu_{G'}/\nu_{G''}$ is roughly constant, with $\nu_{G'}/\nu_{G''} \simeq 2$. {Such a ratio of about 2  has been reported in other soft glassy materials such as carbopol microgels \cite{Migliozzi.2020}, fumed silica grease \cite{Zakani.2018}, or colloid-polymers mixtures{\cite{Koumakis.2012,Truzzolillo.2013}}, and is linked to the way the relaxation time of the sample changes under large strains. The specific ratio $\nu_{G'}/\nu_{G''} = 2$ has been interpreted through simple arguments based upon a Maxwell model, and derived theoretically from Mode-Coupling Theory~\cite{Miyazaki:2006,Wyss:2007}. Here,} for $\gpp>\gpc$, the ratio continuously increases up to $\nu_{G'}/\nu_{G''} \simeq 3$ for the largest explored shear rates, which shows that shear rejuvenation of large intensity yields {a microstructure with a strikingly different evolution of its relaxation time across} the yielding transition. 

\subsection{Impact of flow cessation following  shear rejuvenation on the gel viscoelastic properties} 
\label{FlowCessation}

In all the experiments reported above, the 600~s shear-rejuvenation step was performed at a fixed intensity $\gpp$, and followed by an abrupt flow cessation by imposing $\dot \gamma=0$~s$^{-1}$. In the present section, we explore the impact of flow cessation when performed over a finite duration $\Delta t$ by means of a decreasing ramp of shear stress. In practice, after 600~s of shear rejuvenation at $\gpp=500$~s$^{-1}$, the flow is switched to stress-controlled by imposing $\sigma=\sigmap\simeq 50$~Pa, the latest stress value recorded at the end of the 600~s rate-controlled period. The stress is then swept down from $\sigma_{p}$ to $0$~Pa by discrete steps of amplitude 1 Pa over a total duration $\Delta t$ ranging from $\Delta t=0$~s to $\Delta t\simeq 13000$~s

As shown in the previous two sections, shear rejuvenation at $\gpp>\gpc$ followed by an abrupt flow cessation yields a gel of storage modulus $G'\simeq 500$~Pa. When the flow cessation is performed over a finite duration of increasing value, the storage modulus increases up to a value $G' \simeq 1000$~Pa that is comparable to that obtained after shear rejuvenation at $\gpp<\gpc$ followed by an abrupt flow cessation [see black arrow in Fig~\ref{G_gamma}(b)]. Concomitantly, the loss factor, which is a decreasing function of frequency in the reference case of shear rejuvenation at $\gpp>\gpc$ followed by an abrupt flow cessation, becomes an increasing function of frequency, i.e., $n<0$ [see black arrow in Fig~\ref{G_gamma}(c)].
A similar trend is visible on the nonlinear properties of the boehmite gel when increasing the duration of the flow cessation: the strain $\gamma_0$ at the onset of the nonlinear regime and the yield strain $\gy$ both decrease for increasing $\Delta t$ [see black arrow in Figs.~\ref{sigma_gamma}(b) and \ref{sigma_gamma}(d)]. Moreover, the exponent $\nu_{G'}$ of the power-law decrease of $G'(\gamma)$ beyond the yield point, and the ratio $\nu_{G'}/\nu_{G''}$ both decrease to values comparable to those obtained by shear rejuvenation at $\gpp<\gpc$ followed by an abrupt flow cessation [see black arrow in Figs.~\ref{sigma_gamma}(e) and \ref{sigma_gamma}(f)].

To conclude, varying the duration of the flow cessation following shear rejuvenation at $\gpp>\gpc$ allows us to tune continuously the viscoelastic properties of boehmite gels. Moreover, slow enough flow cessation following shear rejuvenation at $\gpp>\gpc$ provides the boehmite gel with the same properties as shear rejuvenation at $\gpp<\gpc$ followed by an abrupt flow cessation (see Fig.~\ref{supfig1} in Appendix~\ref{Appendix1}). This result shows that the key parameter controlling the terminal properties of the gel at rest is not the largest shear rate imposed to the gel, but rather the period of time during which the sample was sheared at $\dot \gamma <\gpc$. 

\section{Discussion}

Let us now summarize and discuss the results of the present study. The primary sol-gel transition of a boehmite suspension leads to the aggregation of colloidal particles {with a typical size of 10~nm}, into aggregates of radius about 100~nm. These aggregates are unbreakable under shear and act as the new building blocks for any subsequent gelation. The primary gel is therefore constituted of a space-spanning percolated network of clusters of such aggregates. Under external shear the gel yields and turns into a suspension of these clusters, which reorganize under flow and reassemble upon flow cessation, through a secondary gelation. Such a secondary gelation yields consistent and reproducible results that can be repeated numerous times on the same sample provided that it does not suffer from solvent evaporation. 
Here we have shown that the terminal properties of the gel depend on the exact value of the shear {rate} $\gpp$ applied during the shear-rejuvenation step. We have identified a critical shear rate $\gpc$ that delineates two different macroscopic viscoelastic properties upon flow cessation.  {As shown in Appendix~\ref{Appendix_BC} where we use a rough shearing geometry, this phenomenology is a robust property of boehmite gels. Thus, our results} suggest that shear rejuvenation leads to two different microstructures {that we now proceed to tentatively describe} on both sides of $\gpc$. 

\begin{figure}[t!]
\includegraphics[scale=0.43,angle=180]{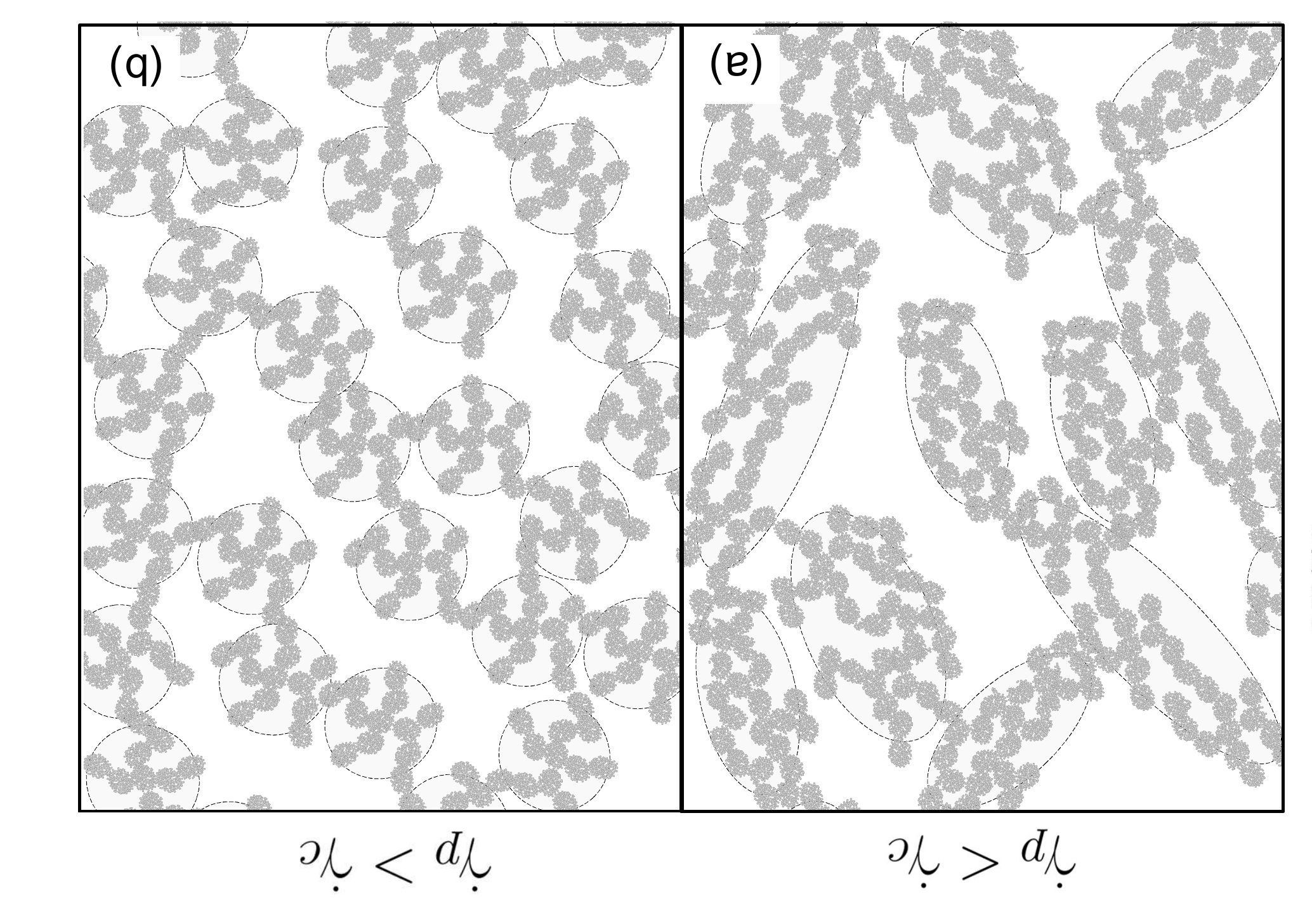}
\caption{\label{structure} Sketch of the gel network microstructure obtained after a shear-rejuvenation step at (a)  $\gpp<\gpc$ yielding large, dense, {and anisotropic} clusters, and at (b) $\gpp>\gpc$, yielding smaller, {isotropic} clusters forming a more open network.} 
\end{figure}

On the one hand, a boehmite gel obtained by an abrupt flow cessation following shear rejuvenation at $\gpp> \gpc$ is characterized by a storage modulus $G' \simeq 500$~Pa comparable to that of the primary gel, and by a viscoelastic spectrum where both the storage and loss moduli are increasing functions of frequency. Moreover, the loss factor $\tan \delta$ also increases with frequency in the same way as the primary gel. Such a viscoelastic spectrum is typical of a \textit{weak colloidal gel} \cite{Trappe:2000,Prasad:2003}, in which the fastest relaxation modes (associated with high frequencies) dominate \cite{Winter:2013}. These observations strongly suggest that the sample is fully fluidized for $\gpp>\gpc$, and that the microstructure that reforms upon flow cessation is similar to that of the primary gel, i.e., a percolated network of small clusters of a few hundred nanometers {[see sketch in Fig.~\ref{structure}(b)]}.

On the other hand, a boehmite gel obtained by an abrupt flow cessation following shear rejuvenation at $\gpp < \gpc$ is characterized by a storage modulus larger than that of the primary gel by a factor of up to 3. Moreover, the viscoelastic spectrum involves a storage modulus that is an increasing function of frequency, whereas the loss modulus is independent or a mildly decreasing function of frequency, similar to aging systems including gels \cite{Aime:2018,Mills:2021} and soft glasses \cite{Purnomo:2008}. {Within the studied frequency range, the} loss factor $\tan \delta$ decreases with increasing frequency, {i.e., the slowest relaxation modes dominate,} which is reminiscent of a glass-like viscoelastic spectrum{\cite{Winter:2013,DelGado.2004,Michele.2011}}. This suggests that the boehmite gel sheared at $\gpp<\gpc$ displays a microstructure composed of clusters {larger} than those constituting the microstructure of the primary gel. 
Such a conclusion is supported by previous experimental and numerical observations that moderate shear during the sol-gel transition of a colloidal suspension favors the growth of denser and thus stronger aggregates \cite{Becker:2009,TuLieu:2016}. {Yet, our observation that the elasticity of boehmite gels is enhanced for $\gpp<\gpc$ strikingly contrasts with the results of Ref.~\cite{Koumakis:2015} on gels of spherical colloidal particles formed through depletion attraction. There, it was shown that preshear at ``low'' rates yielded weaker gels with reduced elasticity due to a largely heterogeneous microstructure inherited from that acquired during shear. To resolve this apparent contradiction, we propose that the shear-assisted growth of the clusters of boehmite unbreakable aggregates results in some \textit{structural anisotropy} within the clusters [see sketch in Fig.~\ref{structure}(a)].}

{Shear-induced anisotropy has indeed been reported many times both in 2D colloidal assemblies through direct visualization \cite{Hoekstra:2003,Masschaele:2011} and in 3D colloidal gels through small-angle light and x-ray scattering \cite{Varadan:2001,Hoekstra:2005}. Very recently, Kao {\it et al.}~\cite{Kao.2021} have demonstrated that, for a given volume fraction in aggregating particles, increasing the particle anisotropy yields gels with a lower fractal dimension and enhanced elastic properties up to a factor 10. In our case, the shape anisotropy would affect the clusters that form under shear, while the building blocks, i.e., the unbreakable aggregates formed upon the primary gelation, remain unchanged. Although more structural investigations are obviously required to confirm such a scenario, we believe that shear-induced anisotropy could easily account for an increase by a factor of 3 in the gel elastic modulus obtained upon cessation of shear at $\gpp<\gpc$. Note also that} an indirect proof of such anisotropy could lie in the decrease of the yield strain for decreasing shear rejuvenation intensity [Fig.~\ref{sigma_gamma}(d)]. Indeed, the anisotropy encoded in the gel microstructure by a continuous shear is expected to strengthen the material along that specific direction of shear \cite{Larson:2019,Wei:2019}. Moreover, a more anisotropic microstructure should also result in a lower resistance of the gel to a change in the shear direction \cite{Grenard:2014}. Since the yield strain is determined by a strain amplitude sweep during which the flow direction changes periodically, we may interpret the decreasing trend seen in $\gy$ upon decreasing $\gpp$ as the signature of an increasing level of anisotropy. 

In that framework, we {may further} propose a microscopic origin for the critical shear rate $\gpc$. Assuming that a gel rejuvenated at $\gpp<\gpc$ displays a microstructure involving clusters made of several unbreakable aggregates of the primary gel, the transition at $\gpc = 30$~s$^{-1}$ corresponds to a critical Mason number ${\rm Mn_c} = 6\pi \eta_s \Rcl^2 \gpc/\mathscr{F}$, where $\eta_s=1$~mPa.s is the solvent viscosity, here water, {$\Rcl\simeq 150$--600~nm} the typical cluster radius, and $\mathscr{F}$ the interaction force between two aggregates in a cluster. 
We may further estimate $\mathscr{F}$ as the van der Waals force between two aggregates of radius $\Rag$, namely $\mathscr{F}  = - A_{H} \Rag/(24 h^{2})$, with $A_{H}$ the Hamaker constant and $h$ the typical distance between the two aggregates~\cite{Hamaker.1937}. This leads to ${\rm Mn_c} = 144\pi \eta_s h^2 \Rcl^2 \gpc/(A_H \Rag)$. Taking $A_{H} = 5.2 kT$ and $h\simeq 10$~nm for boehmite particles~\cite{Nakouzi.2018,Krzysko.2020}, we get ${\rm Mn_c} =$ {0.03-0.5, which is compatible} with a transition controlled by the Mason number \cite{Jamali:2020}.

In other words, as pictured in Fig.~\ref{structure}, shear rejuvenation at $\gpp>\gpc$ leads to the complete separation of all clusters into unbreakable aggregates that subsequently percolate at rest into an isotropic microstructure similar to that of the primary gel. For $\gpp<\gpc$, however, moderate shear allows for the growth of larger and denser clusters {of anisotropic shape}. The microstructure of a gel {exposed to low shear rates is shaped by the external shear, thus bearing some memory of the rejuvenation step through the anisotropy of the microstructure}. 
In this context, stronger gels are obtained when the sample is sheared for a sufficient period of time at $\gpp<\gpc$, which suggests that ($i$) shear rejuvenation at $\gpp<\gpc$ followed by an abrupt flow cessation, and ($ii)$ shear rejuvenation at $\gpp>\gpc$ followed by a slow decreasing ramp of stress should lead to similar microstructures and mechanical properties, in agreement with the results discussed in section~\ref{FlowCessation}.
 
The above scenario is mostly based on the shear-induced {anisotropy of clusters that form a percolated network}. As such, this scenario should neither depend on the boehmite concentration, nor on the acid concentration. To check this hypothesis, we have repeated the measurements reported in sections~\ref{LinearVisco} and \ref{NonlinearVisco} for four different gel compositions. The results summarized in Figs.~\ref{comparaison_boehmite} and \ref{comparaison_acid} in Appendix~\ref{GelChemicalCompo} confirm the robustness of the phenomenology reported in the main text and that the critical shear rate $\gpc$ does not change over a broad range of boehmite and acid concentrations.

{Finally,} our results on boehmite gels show some apparent similarities with previous studies on shear-induced flocculation in colloidal suspensions destabilized by salt. For instance, a critical shear rate $\gpc \simeq 100$~s$^{-1}$ was previously identified, while monitoring the aggregate size in flocculating suspensions of latex particles under shear in a Taylor-Couette cell \cite{Selomulya:2002}. For $\dot \gamma>\gpc$, the aggregate size increases monotonically in time to reach a steady-state value, whereas for $\dot \gamma <\gpc$, the aggregates size displays an overshoot, characterized at long times by a substantial decrease at constant mass, hence corresponding to a restructuring and a densification of the aggregates. As reviewed in Ref.~\cite{Bubakova:2013}, similar results were reported in different geometries, e.g., a mixing tank equipped with various types of impellers \cite{Spicer:1996,Selomulya:2001}, and for non-Brownian particles, e.g., precipitated calcium carbonate \cite{Antunes:2010,Sang:2012}. These reports from the literature indicate that $\gpc \sim 50-100$~s$^{-1}$ for a broad variety of systems and experimental configurations. This suggests that the restructuring of the aggregates under shear in the limit of low shear {rate}, i.e., low Peclet number, is generically {linked to} the strengthening of the elastic properties observed in gels rejuvenated at $\gpp<\gpc$. Nonetheless, the comparison between the literature on shear-induced flocculation and our results cannot be pushed further for none of these previous studies have characterized the rheological properties of the gels that form upon flow cessation after shear rejuvenation below or above $\gpc$. Moreover, previous cluster size measurements were performed on fully dispersed suspensions, whereas our experiments are performed on boehmite gels in their solid-like state. Therefore, comparisons with previous studies should be considered with caution. In order to fully unveil the restructuring scenario below $\gpc$, future experiments will focus on ($i$) measuring the flow profiles during shear rejuvenation, in order to determine whether the flow remains spatially homogeneous or whether it displays heterogeneities such as shear bands, wall slip or fractures, and ($ii$) characterizing the microstructural differences, {including the anisotropy,} induced by the shear history thanks to small-angle light and x-ray scattering measurements.  
 
\begin{acknowledgments}
The authors kindly acknowledge fruitful discussions with Catherine Barentin, Eric Freyssingeas, Thomas Gibaud, Eric L{\'e}colier, {François Li{\'e}nard} and Wilbert Smit. {We also thank the anonymous reviewers for constructive comments on our manuscript.}
\end{acknowledgments}

\appendix 

\section{{Rheological hysteresis}} \label{Appendix_flow_curve}

\begin{figure}[!h]
\includegraphics[scale=0.64]{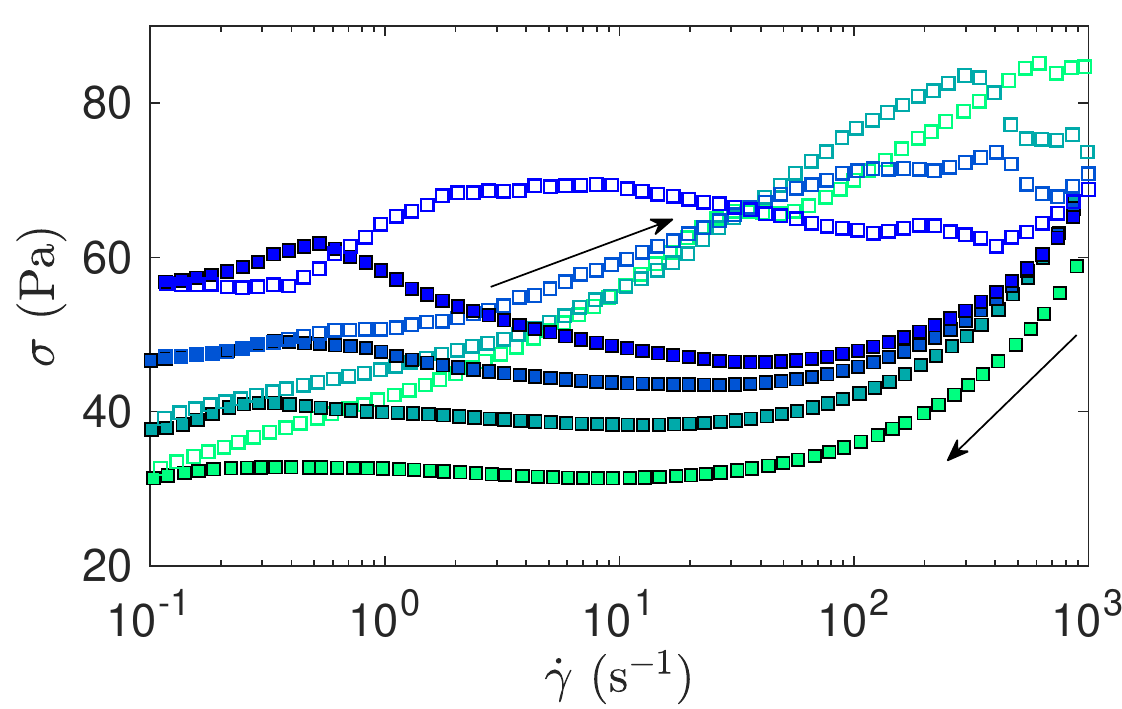}
\caption{\label{fig:flow_curve_cisaillement} Flow curve, shear stress $\sigma$ vs shear rate $\gp$, measured upon imposing a downward sweep in shear rate (full symbols) followed by an upward sweep (empty symbols). The waiting time per point is $\Delta t= 3$ (\textcolor{color10}{$\blacksquare$}), 7.5 (\textcolor{color11}{$\blacksquare$}), 30 (\textcolor{color12}{$\blacksquare$}), and 150~s per point (\textcolor{color13}{$\blacksquare$}). Measurements are performed in a smooth cylindrical Couette geometry (height 58~mm, rotating inner cylinder of radius 24~mm, fixed outer cylinder of radius 25~mm, gap 1~mm).}
\end{figure}

{To illustrate the thixotropy of boehmite gels, downward sweeps in shear rate followed by upward sweeps were performed with different waiting times per point, ranging from 3~s to 150~s per point. As shown in Fig.~\ref{fig:flow_curve_cisaillement}, for all sweep rates, the flow curves show a large hysteresis, which extends up to shear rates way above 10--100~s$^{-1}$, where one usually expects the sample to be fully fluidized \cite{Divoux:2013,DaCruz:2002}. These observations confirm that boehmite gels  are  thixotropic,  and  that  one can expect strong memory effects, as  evidenced by the results discussed in the main text.}

\section{Power-law decay of the viscoelastic modulus beyond the yield point} \label{Appendix2}

\begin{figure}[!h]
\includegraphics[scale=0.72]{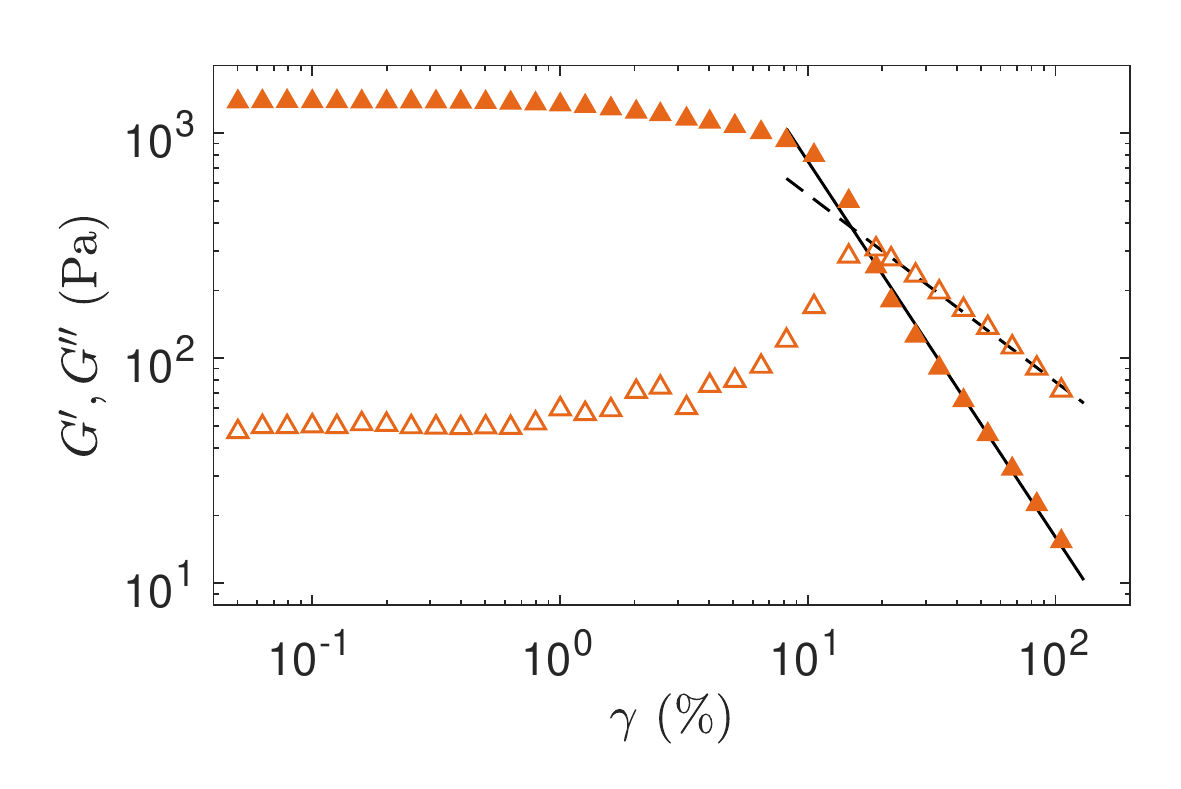}
\caption{\label{supfig2} Storage modulus $G'$ and loss modulus $G''$ vs strain amplitude $\gamma$ during a strain sweep performed at frequency $f=1$~Hz on a boehmite gel following shear rejuvenation at $\gpp=$~2~s$^{-1}$ for 600~s, and a rest period of 3000~s. The continuous and dashed lines respectively correspond to the best power-law fits of $G'$ and $G''$ for $\gamma \geq \gy$.}
\end{figure}

The method used to fit the decay of the storage and loss moduli beyond the yield point defined by $\gy$ is illustrated in Fig.~\ref{supfig2}. Both $G'$ and $G''$ are fitted for $\gamma \geq \gy$ with power laws whose exponents $n_{G'}$ and $n_{G''}$ respectively, are reported in Fig.~\ref{sigma_gamma}(e) in the main text. Note that the exact range of strain fitted to determine these exponents may vary by one or two points between data sets. 

\section{Evolution of the loss factor depending on the flow cessation} \label{Appendix1}

\begin{figure}[!ht]
\includegraphics[scale=0.72]{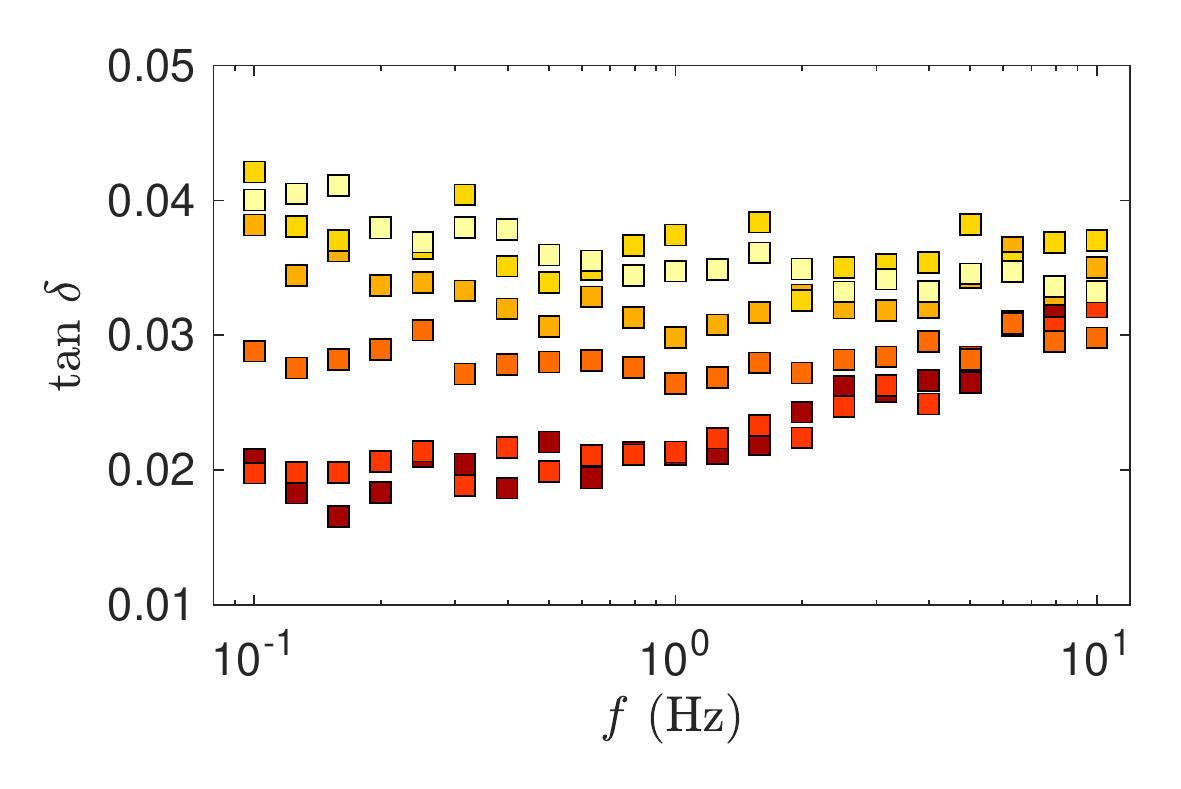}
\caption{\label{supfig1}Frequency dependence of the loss factor $\tan \delta$ depending on the rate of flow cessation from a 600~s shear rejuvenation at $\gpp=500$~s$^{-1}$. The flow is stopped over a finite duration $\Delta t$ using a linear decreasing ramp of shear stress. The duration of the ramp is $\Delta t= 0$ (\textcolor{colorr}{$\blacksquare$}), 54 (\textcolor{color6}{$\blacksquare$}), 270 (\textcolor{color7}{$\blacksquare$}), 540 (\textcolor{color8}{$\blacksquare$}), 3240 (\textcolor{color9}{$\blacksquare$}), and 12,960~s (\textcolor{colorly}{$\blacksquare$}). Viscoelastic spectra measured with a strain amplitude $\gamma=0.1$~\%.}
\end{figure}

{Figure~\ref{supfig1} illustrates the impact of the rate of flow cessation on the frequency dependence of the loss factor $\tan \delta$. Abrupt flow cessation from $\gpp>\gpc$ yields a gel such that $\tan \delta$ increases with the frequency $f$. Decreasing the rate of flow cessation yields gels whose loss factor is less dependent on $f$. Finally, gels obtained after a slow flow cessation show a loss factor that is a decreasing function of $f$ and resembles that of gels prepared through an abrupt flow cessation from $\gpp<\gpc$ [compare with Fig.~\ref{frequency_sweep}(c) in the main text].}

\section{{Impact of boundary conditions}} \label{Appendix_BC}

{To address quantitatively the impact of boundary conditions on the main results of the present work, the shear protocol reported in the main text has been repeated using a rough rotor (surface roughness $\delta \simeq 40~\mu$m obtained by gluing sandpaper) instead of a smooth rotor.}

\begin{figure}[!h]
\includegraphics[scale=0.57]{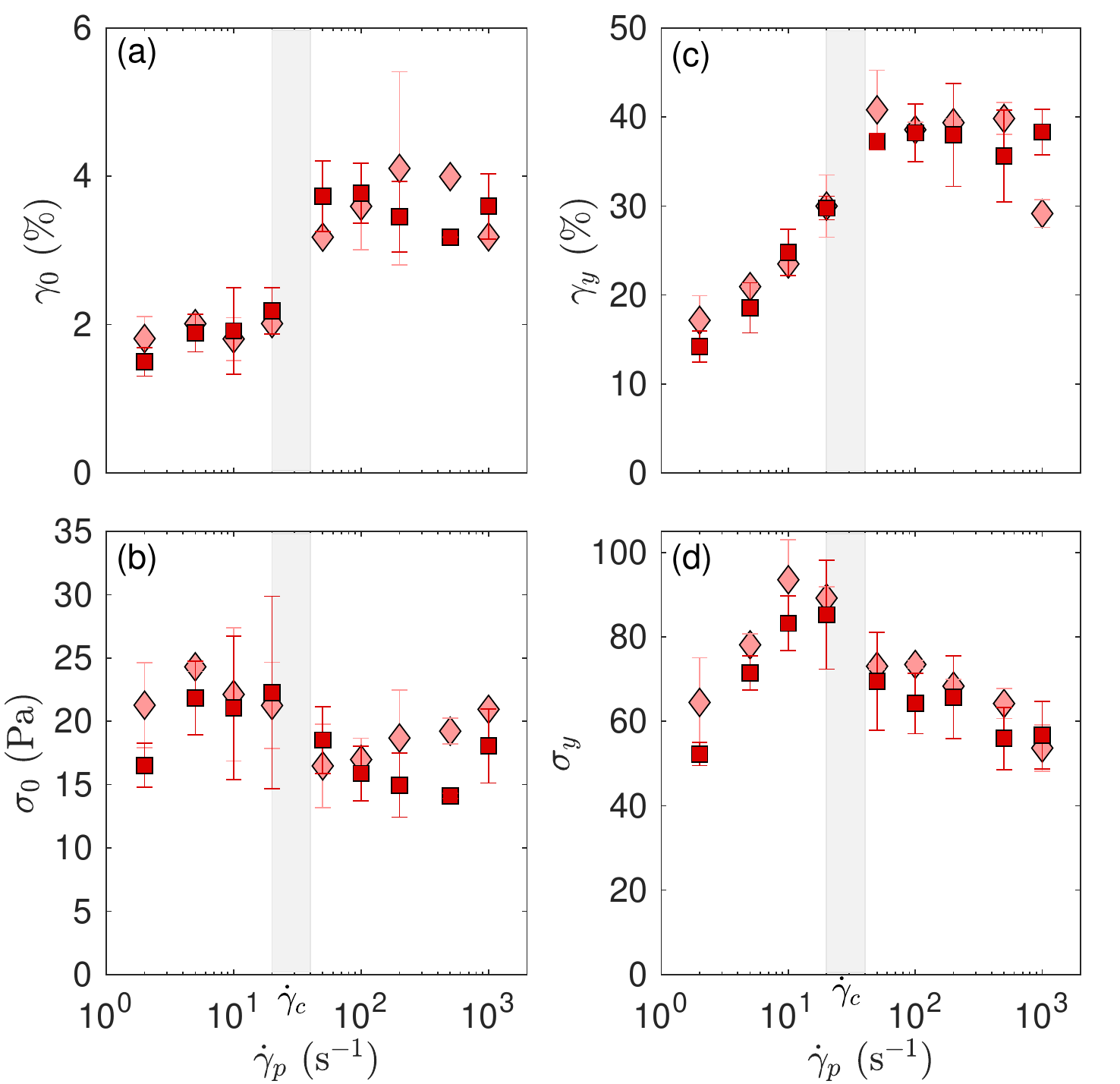}
\caption{\label{fig:sigma_gamma_geometry} (a) Strain amplitude $\gamma_{0}$ at the onset of the nonlinear regime and (b) corresponding stress amplitude $\sigma_{0}$ vs the shear rate $\gpp$ applied during the shear-rejuvenation step. (c) Yield strain $\gy$ and (d) yield stress $\sy$ vs $\gpp$ for a smooth  rotor (\textcolor{colorr}{$\blacksquare$}) and a rough rotor (radius of 23.9 mm)~(\textcolor{colorlr}{$\blacklozenge$}).}
\end{figure}

\begin{figure}[!h]
    \centering
    \includegraphics[scale=0.72]{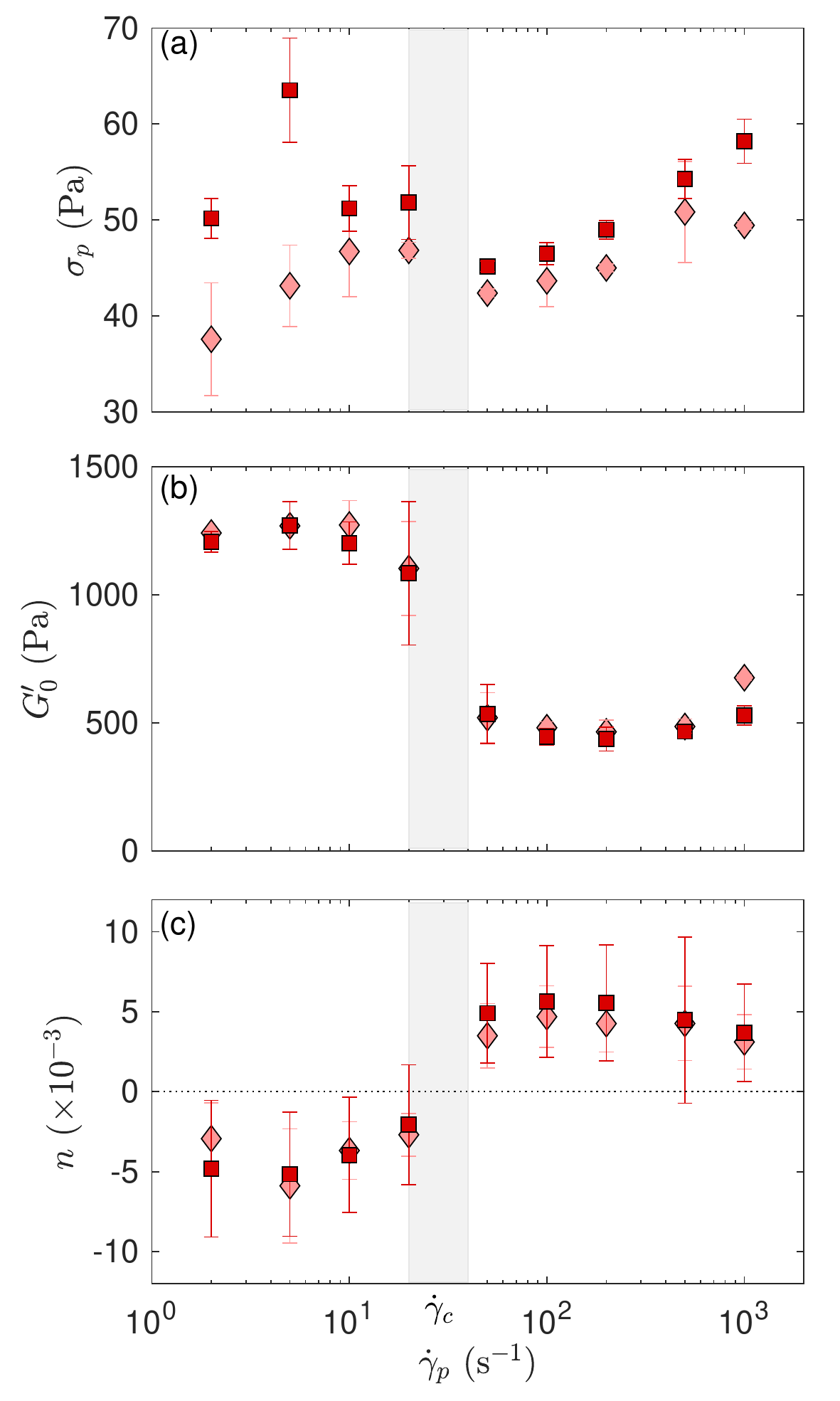}
    \caption{\label{fig:G_gamma_geometry}Dependence on the shear rate $\gpp$ used for shear rejuvenation of (a) the stress $\sigma_{p}$ measured at the end of the shear-rejuvenation step, (b) the storage modulus $G'$ measured after 3000~s of rest period following the shear-rejuvenation step, and (c) the slope $n$ of the linear regression of the loss factor $\tan\delta$ versus $\log f$.  Same symbols and color code as in Fig. \ref{fig:sigma_gamma_geometry}.}
\end{figure}

{Figures~\ref{fig:sigma_gamma_geometry} and \ref{fig:G_gamma_geometry} show a direct comparison of the linear and non-linear viscoelastic properties of the same boehmite gel measured with either a smooth rotor (red squares) or a rough rotor (pink diamonds). Up to experimental uncertainty, the results are the same over the entire range of shear rates $\gp$ applied during the rejuvenation step, except maybe for $\sigmap$ at low $\gp$. These experimental results show that, if present, wall slip does not affect the conclusions presented in the manuscript, and that the critical shear rate $\dot \gamma_c$, which separates the two different rheological regimes, is a robust property of boehmite gels.}

\section{Impact of the gel chemical composition}
\label{GelChemicalCompo}

\begin{figure}
\includegraphics[scale=0.55]{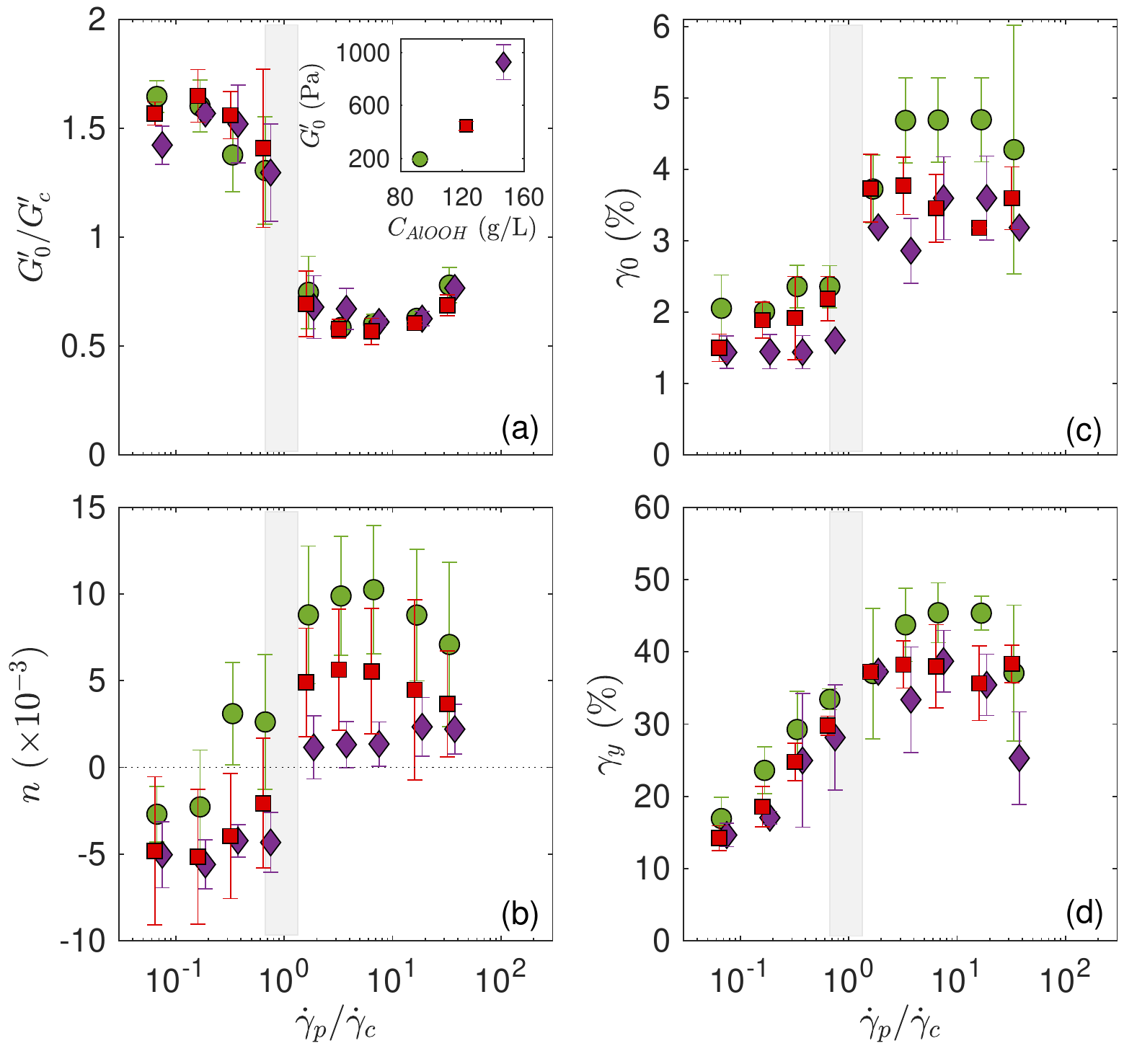}
\caption{\label{comparaison_boehmite} Impact of boehmite concentration. (a) Normalized storage modulus $G'_0$/$G'_{\rm c}$ measured 3000~s after the abrupt cessation of a 600~s shear rejuvenation performed at $\gpp$, (b) slope of the linear regression of the loss factor versus $\log f$, (c) strain $\gamma_0$ at the onset of the nonlinear regime, and (d) yield strain $\gy$ as a function of $\gpp$/$\dot{\gamma}_{c}$, where $\gpc$ is the critical shear rate (see Table~\ref{tab:table1}). Symbols stand for gels with different boehmite concentrations and the same nitric acid concentration [$\text{B}=93,\text{A}=14$] (\textcolor{colorg}{$\bullet$}), [$\text{B}=123,\text{A}=14$] (\textcolor{colorr}{$\blacksquare$}) and [$\text{B}=147,\text{A}=14$] (\textcolor{colorp}{$\blacklozenge$}). In (a), $G'_c$ is the value of $G'_0$ extrapolated at $\gpp=\gpc$ and the inset shows $G^{\prime}_{0}$ vs the boehmite concentration.
The grey rectangle highlights the transition region around $\gpc$.}
\end{figure}

\begin{figure}
\includegraphics[scale=0.55]{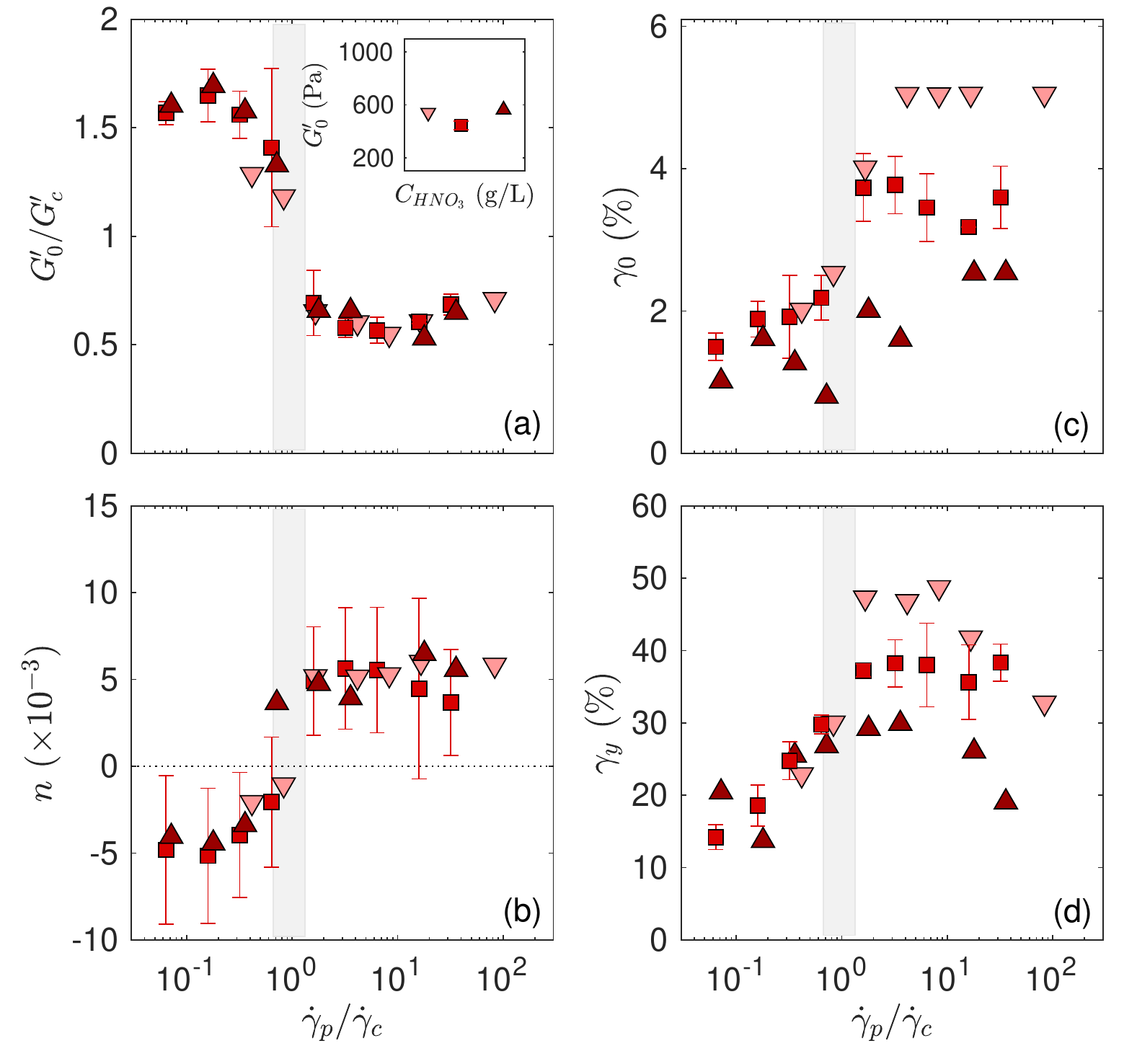}
\caption{\label{comparaison_acid} Impact of the acid concentration. (a) Normalized storage modulus $G'_0$/$G'_{\rm c}$ measured 3000~s after the abrupt cessation of a 600~s shear rejuvenation performed at $\gpp$, (b) slope of the linear regression of the loss factor versus $\log f$, (c) strain $\gamma_0$ at the onset of the nonlinear regime, and (d) yield strain $\gy$ as a function of $\gpp$/$\dot{\gamma}_{c}$, where $\gpc$ is the critical shear rate (see Table~\ref{tab:table1}). Symbols stand for gels with different acid concentrations and the same boehmite concentration [$\text{B}=123,\text{A}=12$] (\textcolor{colorlr}{$\blacktriangledown$}), [$\text{B}=123,\text{A}=14$] (\textcolor{colorr}{$\blacksquare$}) and [$\text{B}=123,\text{A}=17$] (\textcolor{colorvdr}{$\blacktriangle$}). In (a), $G'_c$ is the value of $G'_0$ extrapolated at $\gpp=\gpc$ and the inset shows $G^{\prime}_{0}$ vs the acid concentration.  
The grey rectangle highlights the transition region around $\gpc$.}
\end{figure}

On top of the sample with 123~g.L$^{-1}$ of boehmite and 14~g.L$^{-1}$ of nitric acid reported in the main text, we have investigated the four supplementary compositions gathered in Table~\ref{tab:table1} by following the same shear rejuvenation and characterization protocols.
The impact of boehmite and acid concentrations on the various linear and nonlinear rheological observables is reported in Figs.~\ref{comparaison_boehmite} and \ref{comparaison_acid}, respectively. For the sake of clarity, the storage modulus $G'_0$ is normalized by $G'_c$, defined as the value of $G'_0$ extrapolated at $\gpp=\gpc$, while the shear rate $\gpp$ used for shear rejuvenation is normalized by $\gpc$.

The storage modulus $G'_0$ of boehmite gels increases steeply with the concentration of boehmite [inset in Fig.~\ref{comparaison_boehmite}(a)], whereas it is roughly insensitive to a change of the acid concentration for a given boehmite content [inset in Fig.~\ref{comparaison_acid}(a)]. Yet, for all gel compositions, the storage modulus $G'_0$ can be rescaled onto a master curve that shows two distinct values on both sides of a critical shear rate $\gpc$. Shear rejuvenation at $\gpp<\gpc$ always yields a higher storage modulus than shear rejuvenation at $\gpp>\gpc$ [Figs.~\ref{comparaison_boehmite}(a) and \ref{comparaison_acid}(a)]. Moreover, $\gpc$ appears to be independent of the boehmite and acid contents over the range of compositions under study [see Table~\ref{tab:table1}]. These results confirm the robustness of the conclusions presented in the main text in Fig.~\ref{G_gamma}(b).  

As for the linear viscoelastic spectrum, Fig.~\ref{comparaison_boehmite}(b) and \ref{comparaison_acid}(b) show that the slope $n$ of the linear regression of the loss factor versus $\log f$ is negative for $\gpp < \gpc$, whereas $n>0$ for $\gpp > \gpc$. Therefore, for all the compositions investigated here, boehmite gels display two different relaxation spectra on both sides of $\gpc$, and two different microstructures. Note that the lower boehmite concentrations yield larger values of $n$, which is otherwise insensitive to the acid concentration. 

Finally, both the strain $\gamma_0$ at the onset of the nonlinear regime and the yield strain $\gy$ display a dual behavior on both sides of $\gpc$ for all the compositions investigated, in agreement with Figs.~\ref{sigma_gamma}(a) and \ref{sigma_gamma}(c) in the main text. While $\gamma_0$ and $\gy$ are poorly sensitive to the boehmite content [Figs.~\ref{comparaison_boehmite}(c) and \ref{comparaison_boehmite}(d)], they display an overall increasing trend for decreasing acid content for $\gpp>\gpc$ [Figs.~\ref{comparaison_acid}(c) and \ref{comparaison_acid}(d)]. This points to a dependence of the gel microstructure with the concentration of nitric acid that remains to be elucidated through structural measurements.

\begin{table*}[ht!]
\caption{\label{tab:table1}}
\begin{ruledtabular}
\begin{tabular}{cccccccc}
Gel & Boehmite concentration (g.L$^{-1}$) & Acid concentration (g.L$^{-1}$) &$\dot{\gamma}_{c}$ (s$^{-1}$) & $G'_{\rm c}$ (Pa)\\
\hline
[$\text{B}=93,\text{A}=14$] & 93 & 14 & 30 $\pm$ 10  & $330\pm 25$\\

[$\text{B}=123,\text{A}=14$] & 123 & 14 & 30 $\pm$ 10 & $760\pm 60$\\

[$\text{B}=147,\text{A}=14$] & 147 & 14 & 30 $\pm$ 10  & $1380\pm 70$\\

[$\text{B}=123,\text{A}=13$] & 123 & 12 & 10$\pm$ 10 & $990\pm 60$\\

[$\text{B}=123,\text{A}=18$] & 123 & 17 & $30\pm10$ & $860\pm 60$\\
\end{tabular}
\end{ruledtabular}
\end{table*}

\newpage

\section*{References}

%

\end{document}